\documentclass[journal]{IEEEtran}
\ifCLASSINFOpdf
\else
\fi
\usepackage{amssymb}
\usepackage{amstext}
\usepackage{amsmath}
\usepackage{amsthm}
\usepackage{epsfig}
\usepackage{epstopdf}
\usepackage{calc}
\usepackage{color}
\usepackage{graphicx}
\usepackage{caption}
\usepackage{booktabs}
\usepackage{rotating}
\usepackage{pdflscape}
\usepackage{slashbox}
\usepackage{subfig}

\begin{document}
\title{Nonlinear Model Predictive Control of A Gasoline HCCI Engine Using Extreme Learning Machines}
\author{Vijay Manikandan Janakiraman,
        XuanLong Nguyen,
        and Dennis Assanis % <-this % stops a space
\thanks{Vijay Manikandan Janakiraman is with NASA Ames Research Center (UARC), Moffett Field, CA, USA. He was previously with the Department of Mechanical Engineering, University of Michigan, Ann Arbor, MI, USA e-mail: vijai@umich.edu.}% <-this % stops a space
\thanks{XuanLong Nguyen is with the Department of Statistics, University of Michigan, Ann Arbor, MI, USA.}% <-this % stops a space
\thanks{Dennis Assanis is with the Stony Brook University, NY, USA.}}% <-this % stops a space
\markboth{In Review, Special Issue on Neurodynamic Systems for Optimization and Applications.}%
{Shell \MakeLowercase{\textit{et al.}}: Nonlinear Model Predictive Control of Gasoline HCCI Engines Using Extreme Learning Machines}
\maketitle
%\begin{abstract}
%Homogeneous charge compression ignition (HCCI) is a futuristic combustion technology that operates with a high fuel efficiency and reduced emissions. HCCI combustion is characterized by complex nonlinear dynamics which necessitates a model based control approach for automotive implementation. HCCI engine control is a nonlinear, multi-input multi-output problem with state and actuator constraints which makes the controller design very challenging. Early control designs make use of a first principles based model which involves a long development time and cost associated with expert labor and calibration. In this paper, an effective data based approach is presented using extreme learning machines (ELM) and nonlinear model predictive control (MPC). ELM is used to develop a nonlinear dynamic model of the engine using experimental data and is shown to accurately represent the engine in a compact manner. A design of experiments for nonlinear system identification is employed for efficient data collection. Using the ELM engine models, an MPC based control algorithm with a simplified quadratic program update is derived for real time implementation. The working and effectiveness of the MPC approach has been analyzed on a nonlinear HCCI engine model for tracking multiple reference quantities along with constraints defined by HCCI states, actuators and operational limits.
%\end{abstract}

% \IEEEpeerreviewmaketitle
\begin{abstract}
Homogeneous charge compression ignition (HCCI) is a futuristic combustion technology that operates with a high fuel efficiency and reduced emissions. HCCI combustion is characterized by complex nonlinear dynamics which necessitates a model based control approach for automotive application. HCCI engine control is a nonlinear, multi-input multi-output problem with state and actuator constraints which makes controller design a challenging task. Typical HCCI controllers make use of a first principles based model which involves a long development time and cost associated with expert labor and calibration. In this paper, an alternative approach based on machine learning is presented using extreme learning machines (ELM) and nonlinear model predictive control (MPC). A recurrent ELM is used to learn the nonlinear dynamics of HCCI engine using experimental data and is shown to accurately predict the engine behavior several steps ahead in time, suitable for predictive control. Using the ELM engine models, an MPC based control algorithm with a simplified quadratic program update is derived for real time implementation. The working and effectiveness of the MPC approach has been analyzed on a nonlinear HCCI engine model for tracking multiple reference quantities along with constraints defined by HCCI states, actuators and operational limits.
\end{abstract}


\section{Introduction}\label{intro sec}
In recent years, the requirements on automotive performance, emissions and cost have become increasingly stringent. As a consequence, the auto industry has been continuously introducing efficient mechanisms to meet these demands \cite{auto_ind}. Invariably, such systems introduce additional complexity and associated challenges in design and implementation. Homogeneous charge compression ignition (HCCI) is an advanced combustion technology incorporating several of the recent automotive advancements including variable valve timing, exhaust gas recirculation, intake charge boosting etc \cite{auto_ind}. HCCI engines shifted the spotlight from traditional spark ignited (SI) and compression ignited (CI) engines owing to its ability to reduce emissions and fuel consumption significantly \cite{thring,Aoyama}. However, HCCI poses several challenges for implementation. These include the absence of a direct trigger for combustion, narrow operating range and high sensitivity to ambient conditions amongst others. Control of HCCI combustion is a challenging problem and a model based approach is typically employed \cite{Johansson2010,Ravi2009}, where a control-oriented reduced order model is developed using first principles. Such models involve a long development time and cost associated with expert labor and calibration. To accelerate HCCI implementation on automotive applications, a key requirement is to develop predictive dynamic models quickly that can not only capture the required dynamics for control but also operate under limited resources, for instance, onboard an engine electronic control unit (ECU) whose memory and computation are limited.

In order to enable the complex control strategies, the HCCI engine is equipped with additional sensors such as in-cylinder pressure transducers that can provide valuable operational data. In-cylinder pressure data can give insights into the general combustion behavior of an engine, including efficiency, emissions and stability. Although the primary use is to observe the states of the engine for feedback control \cite{ca50_1,ca50_2}, such sensory information can be processed to develop dynamic models of the system itself. It has been previously shown by the authors that in-cylinder pressure data can be used to develop control oriented models of engine combustion phasing, work output, combustion stability using neural networks \cite{vijay_asoc} and support vector machines \cite{vijay_springer,vijay_ieee}. This article considers controller design for HCCI engine using data based models. Control systems using machine learning models are popular in robotics, autonomous and aerospace systems \cite{aero_ref,robotics_ref} but less popular in the automotive industry. However, with increase in complexity of emerging systems and with advanced sensing capabilities, there is a significant need to identify new venues and evaluate the potential of information based models for automotive control applications. This forms the main motivation of the interdisciplinary research considered in this article.

For the HCCI modeling problem, Extreme Learning Machines (ELM) were selected for their fast operation and approximation capabilities to fit nonlinear systems \cite{4Huang2005}. Also, when ELM is trained with real world experimental data, it approximates the real system and makes no (or minimal) simplifying assumptions about the underlying phenomena. The dynamics of sensors, actuators and other complex processes which are usually overlooked/hard to model using first principles, can be captured using the identification method. In addition, for a system like the combustion engine, prototype hardware is typically available and extensive experimental data can be collected making the data based approach more appropriate. Data based identification is less common for HCCI engines owing to its nonlinear, highly sensitive and unstable behavior. However, it has been shown previously by the authors that artificial neural networks \cite{vijay_asoc} and support vector machines \cite{vijay_springer} are indeed suited for accurately predicting the dynamics of HCCI. However, artificial neural networks (based on backpropagation) have issues with picking up a local minima while support vector machines result in a large number of parameters making it unsuitable for real engine implementations \cite{vijay_springer}. ELMs on the other hand, solve a convex problem resulting in a global optimal solution and simple to be implemented onboard the engine ECU to make predictions.

For the HCCI control problem, a model predictive control (MPC) framework is employed that can operate with black box models \cite{maciej_nn}. In addition, MPC is well suited for handling operation related and hardware related constraints in an elegant manner. Even for simple linear systems, MPC has been shown to perform well against traditional control such as PID \cite{maciej}, LQ control \cite{LQ1,LQ2} etc. Typically, MPC makes use of a mathematical model of the system and solves an optimization problem with the given constraints to achieve an optimal control solution \cite{maciej}. In this paper, a recurrent ELM engine model is used to make predictions which are then used by MPC to make control decisions for tracking a given reference command. Model predictive control has been quite popular in the automotive domain for both spark ignited \cite{mpc_ref5} and compression ignited engines \cite{mpc_ref3,mpc_ref4}. However, the application of MPC to HCCI engines is at an early stage involving physics based models \cite{mpc_ref6,mpc_ref7} and simple identification models \cite{mpc_ref8}. This article advances the MPC application to HCCI engines by demonstrating on a data based non-linear model that could capture more complex behavior of the engine.

The contributions of the article can be summarized as follows. An ELM based system identification framework is developed for modeling the HCCI engine from real experimental data. This include optimal model selection (tuning model hyper-parameters), training and validation of the model to the HCCI engine data. An ELM based MPC framework is not reported in the literature to the best of the authors knowledge. A control oriented model of the gasoline HCCI engine predicting engine work output, combustion phasing, combustion stability in terms of maximum pressure and maximum pressure rise rate etc. is developed using ELM models and validated at several operating conditions. A model predictive control framework based on linearized ELM and a simplified quadratic program update is developed for the HCCI system suitable for real-time application. The article finally identifies the HCCI engine as a novel application domain for demonstration of the above methodologies in combination.

The article is organized as follows. The HCCI engine experimental setup, design of experiments and data collection are summarized in section \ref{expt_sec}. A system identification procedure using ELM is described in section \ref{modeling_sec} where predictive models are developed and validated using HCCI experimental data. A model predictive control algorithm using linearized ELM models is derived in section \ref{mpc_sec} where linearization is done using the structure of ELM to avoid numerical gradient calculations. Finally, the ELM based MPC approach is demonstrated in simulations in section \ref{sim sec}. A fast quadratic programming method is employed using the convex nature of the problem which enables real-time implementation of the controller. Simulation cases have been analyzed to understand the working and effectiveness of the proposed method.

\section{HCCI Engine System And Experimentation}\label{expt_sec}
\begin{figure*}[]
      \centering
      \includegraphics[scale=0.6]{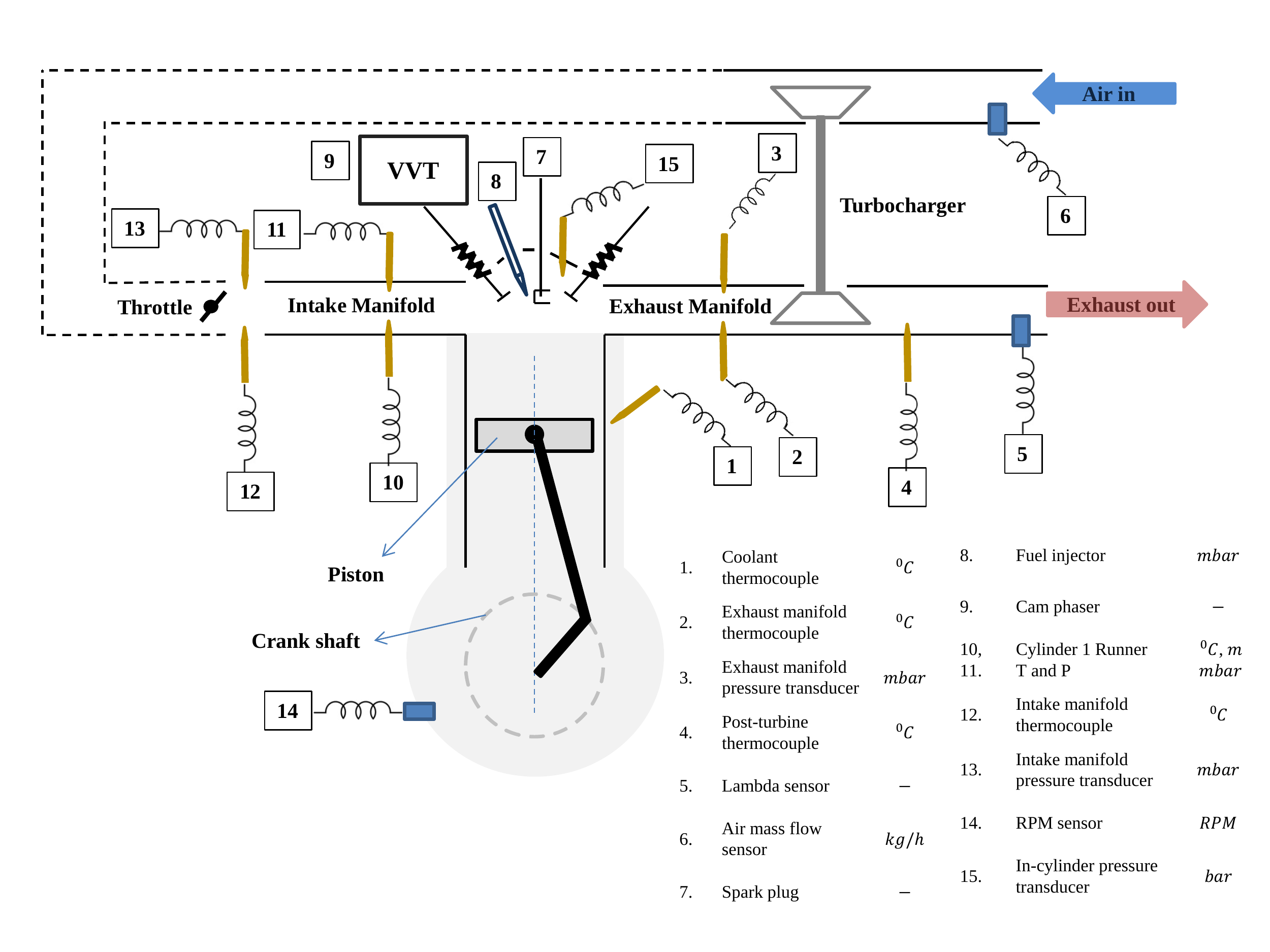}
      \caption{A schematic of the HCCI engine setup and instrumentation (only relevant instrumentation is shown).}
      \label{schematic}
\end{figure*}

In this section, the HCCI engine basics, experimental setup and data collection strategy are described. Majority of this section has been adopted from \cite{vijay_asoc} and included in this article for completeness. The engine (specifications listed in Table \ref{specstable}) is a 4 stroke, inline 4 cylinder gasoline engine with modifications to enable HCCI operation, such as increased compression ratio, direct fuel injection, intake boosting and variable valve timing. A schematic of the experimental setup and instrumentation is shown in Fig. \ref{schematic}. The sensors of the engine include a fast in-cylinder pressure transducer, thermocouples at several locations at the intake and exhaust manifold, cylinder runners etc. The engine is operated with natural aspiration, i.e., all experiments in this work involve no or negligible difference between intake and exhaust manifold pressures.

\begin{table}[hbtp]
\caption{Specifications of the experimental HCCI engine}
\label{specstable}
\footnotesize
\begin{center}
\begin{tabular}[c]{|c|c|}
\hline
Engine Type & 4-stroke In-line\\
\hline
Fuel & Gasoline\\
\hline
Displacement & 2.0 L\\
\hline
Bore/Stroke & 86/86 mm\\
\hline
Compression Ratio & 11.25:1\\
\hline
Injection Type & Direct Injection\\
\hline
 & Variable Valve Timing with \\
&  hydraulic cam phaser having\\
Valvetrain  & 119 degree constant duration \\
 & defined at 0.25mm lift, 3.5mm peak \\
 & lift and 50 degree crank angle \\
 & phasing authority \\
 \hline
HCCI strategy & Exhaust recompression \\
& using negative valve overlap\\
\hline
\end{tabular}
\end{center}
\end{table}

\subsection{HCCI Strategy}
\begin{figure}[]
      \centering
      \includegraphics[scale=0.45]{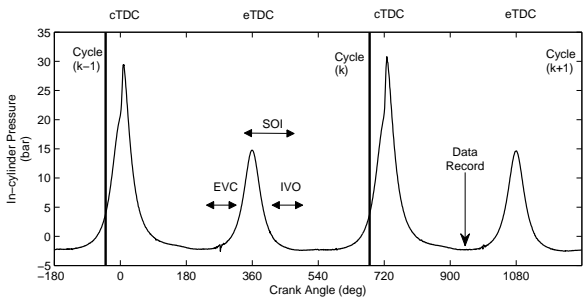}
      \caption{The pressure trace of a HCCI combustion cycle showing cycle definition, actuator ranges of intake valve opening (IVO), exhaust valve closing (EVC), start of injection (SOI). The crank angle at data recording are also shown.}
      \label{cycle defn}
\end{figure}

HCCI operation is achieved using a recompression strategy. The engine is a discrete event system where one combustion cycle (720 deg crank angle) represents an event. A snapshot of the cylinder pressure trace of one combustion cycle along with valve events and fuel injection events are shown in Fig. \ref{cycle defn}. During every combustion cycle, three distinct control actions can be defined - the crank angle at intake valve opening (IVO), crank angle at exhaust valve closing (EVC) and crank angle at start of fuel injection (SOI). The valve events are measured in degrees after exhaust top dead center (deg eTDC) while SOI is measured in degrees after combustion top dead center (deg cTDC). It should be noted that the fuel mass (FM in mg/cyc) is injected in a region between EVC and IVO defined as the negative valve overlap (NVO). The fuel is injected early (during NVO of previous cycle) and given sufficient time to mix with air forming a homogeneous mixture. A large fraction of the exhaust gas from the previous combustion cycle (EGR) is retained to elevate the temperature and hence the reaction rates of the fuel and air mixture. The variable valve timing capability of the engine enables trapping suitable quantities of exhaust gas in the cylinder. The fuel-air mixture is given sufficient time to mix homogeneously and as it is compressed during the compression stroke, auto-ignition occurs initiating combustion. Since the fuel is mixed homogeneously throughout the cylinder, combustion of the fuel-air mixture happens almost instantaneously, releasing heat and pushing the piston down performing work. More details on setup and operation of the HCCI engine can be found in \cite{vijay_asoc,vijay_ieee}.

\subsection{HCCI Inputs and Outputs}
HCCI combustion is dominated by chemical kinetics and is limited by stability constraints such as misfire and knock \cite{misf1,misf2}. To achieve a stable combustion, a right proportion of fuel, air and EGR is required for a given thermal condition of the engine. Further, the combustion must occur at the right instant during the expansion stroke (given by CA50) for high efficiency and low emissions. By varying the valve events, the quantity of EGR can be varied while FM and SOI can control the ignition delay that affects combustion phasing. Thus, the engine operation can be controlled using quantities such as FM, IVO, EVC and SOI. Other important physical variables that influence the performance of HCCI combustion include intake manifold temperature $T_{in}$, intake manifold pressure $P_{in}$, mass flow rate of air at intake $\dot{m}_{in}$, exhaust gas temperature $T_{ex}$, exhaust manifold pressure $P_{ex}$, coolant temperature $T_{c}$, fuel to air ratio (FA) etc.

The engine performance metrics can be defined as follows. The indicated mean effective pressure (IMEP) represents the work output from the engine and is a primary response variable to be controlled. For instance, the driver's power demand can be interpreted as a change in desired IMEP command given to the engine controller. The second and most important performance metric is the combustion phasing indicated by the crank angle at 50\% mass fraction burned (CA50). The CA50 is a key quantity that influences engine performance, efficiency, emissions and stability \cite{ca50_1,ca50_2}. Further, additional quantities that give an indication of the combustion stability such as maximum pressure ($P_{max}$) and maximum pressure rise rate ($R_{max}$) in a combustion cycle are considered performance metrics to be monitored. The equivalent fuel air ratio (EAFR) that indicates if the mixture in the cylinder is fuel rich or fuel lean is another important parameter to be monitored. For HCCI control discussed in this article, the control knobs such as FM, EVC and SOI are considered inputs while performance variables such as IMEP, CA50, $P_{max}$, $R_{max}$, engine torque and EAFR are considered outputs. The IMEP, CA50, $P_{max}$ and $R_{max}$ are calculated from the high speed in-cylinder pressure measurements. For further reading on HCCI combustion and related variables, please refer \cite{hcci_book,vijay_asoc,vijay_ieee}.

\subsection{Experiment Design}
\begin{figure}[]
      \centering
      \includegraphics[scale=0.45]{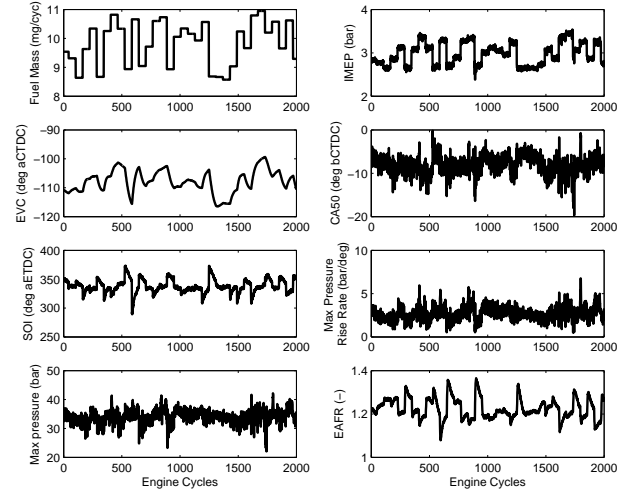}
      \caption{A subset of closed loop experimental data showing the A-PRBS inputs and the measured engine outputs.}
      \label{cl_data fig}
\end{figure}

The goal of the experiments is to obtain dynamic data from the HCCI experimental setup for model development. This task falls into the category of nonlinear system identification \cite{nelles}. In order to obtain sufficiently rich information from the system, the excitation signal must excite the system at all frequencies and amplitudes \cite{prbs}. Persistent excitation cannot be guaranteed for nonlinear systems \cite{prbs} but in practice, amplitude modulated pseudo-random binary sequence (A-PRBS) signals have shown good identification performance \cite{nelles}. Thus an A-PRBS signal exciting the engine at different amplitudes and frequencies is considered in this work. A set of transient experiments is conducted at a constant speed of 1800 RPM and naturally aspirated conditions using a feedback controller developed for HCCI using first principles \cite{jade2012}. An A-PRBS sequence of reference IMEP and CA50 are given as inputs to the feedback controller while the controller calculates the commands of FM, EVC and SOI to be given to the engine. The experiments are conducted and data recorded using specialized engine rapid prototyping hardware. The data is sampled using the AVL Indiset acquisition system where in-cylinder pressure is sensed every crank angle while IMEP, CA50, $P_{max}$ and $R_{max}$ are determined on a per-combustion cycle basis. A subset of the HCCI input and output sensor data can be shown in Fig. \ref{cl_data fig}. The data is processed, scaled and converted to the input format for ELM training. More details on HCCI combustion, experiments and data preprocessing can be found in \cite{vijay_springer,vijay_asoc}.

\section{HCCI Engine Model Development Using Extreme Learning Machines}\label{modeling_sec}
\subsection{Learning HCCI Dynamics using Recurrent Models}
\begin{figure*}[]
      \centering
      \begin{tabular}{cc}
      \subfloat[Series-Parallel architecture for system identification.]{\label{ser-par_pic}\includegraphics[scale=0.35]{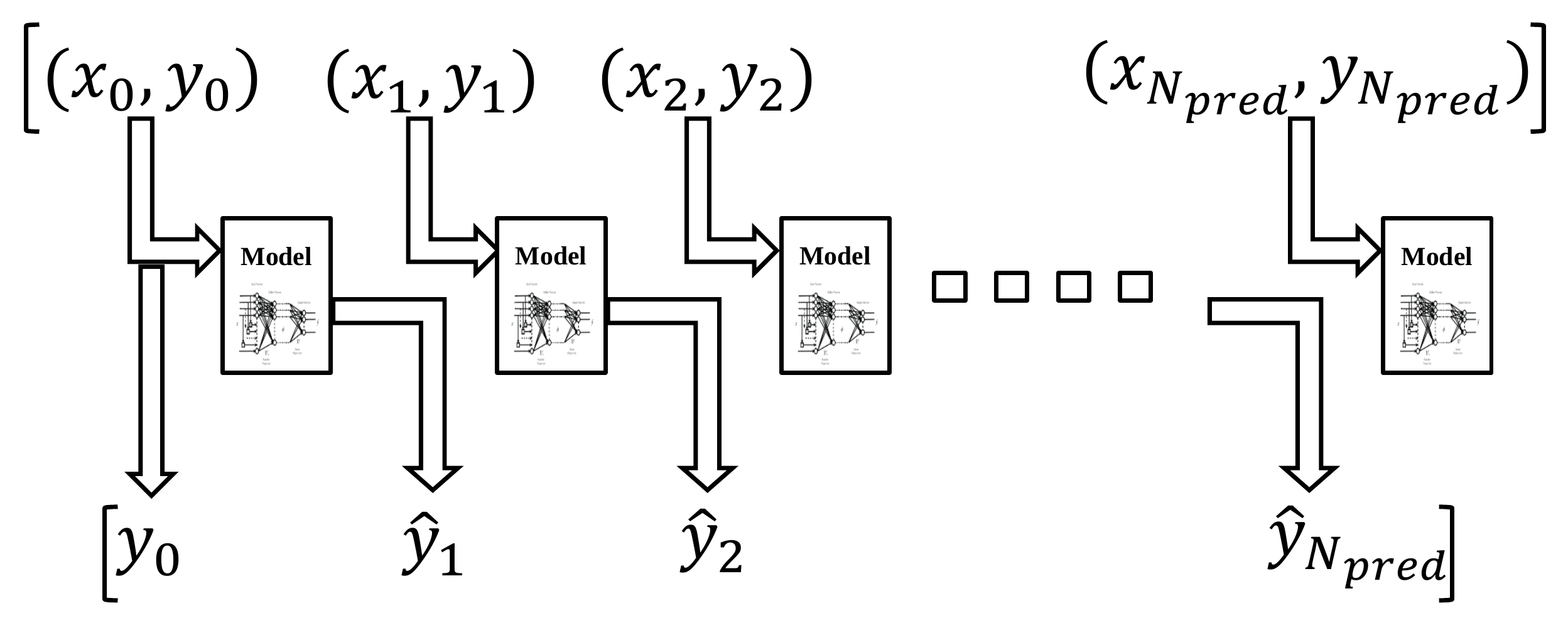}}
      &      \subfloat[Parallel Architecture for system identification.]{\label{par_pic}\includegraphics[scale=0.35]{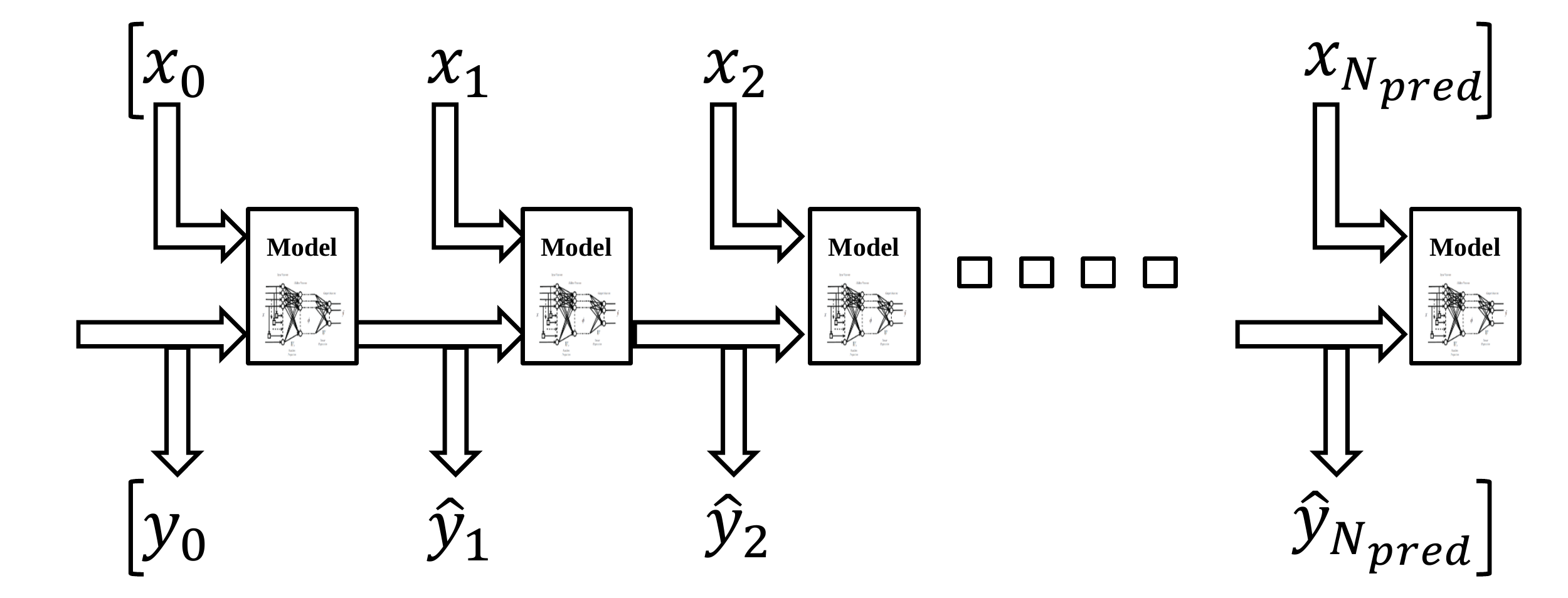}}\\
      \end{tabular}
      \caption{}
      \label{}
\end{figure*}

For the HCCI engine system, both the inputs and the outputs of the engine are available as sensor measurements and naturally, supervised learning methods can be used. The HCCI engine is a nonlinear dynamic system and sensor measurements represent discrete time sequences. A nonlinear auto regressive model with exogenous input (NARX) \cite{nelles} can be considered as follows
\begin{equation}\label{NARX eqn}
y(k)=f_{NARX}[u(k-1),..,u(k-n_u),y(k-1),..,y(k-n_y)],
\end{equation}
where $u(k)\in\mathbb{R}^{u_d}$ and $y(k)\in\mathbb{R}^{y_d}$ represent the inputs and outputs of the system respectively, $k$ represents the discrete time index, $f_{NARX}(.)$ represents the nonlinear function mapping specified by the model, $n_u$, $n_y$ represent the number of past input and output samples required (order of the system) while $u_d$ and $y_d$ represent the dimension of inputs and outputs respectively. Let
\begin{equation}\label{inp feature}
x(k)=[u(k-1),..,u(k-n_u),y(k-1),..,y(k-n_y)]^T
\end{equation}
represent the augmented input vector obtained by appending the input and output measurements from the system. The engine measurement sequence can be converted to the form of training data
\begin{equation}\label{}
\{(x_1,y_1),...,(x_N,y_N)\}\in \big(\mathcal{X},\mathcal{Y}\big)
\end{equation}
where $N$ denotes the number of training samples, $\mathcal{X}$ denotes the space of the input features (Here $\mathcal{X} = \mathbb{R}^{u_dn_u+y_dn_y}$ and $\mathcal{Y} = \mathbb{R}^{y_d}$). The above conversion of system measurements to training data can be used to train a series-parallel model where the past inputs and outputs of the system (concatenated in $x$) is used for one-step ahead predictions (OSAP) i.e., given a set of measurements until time index $k$, the model predicts the output at time $k+1$ (see equation \eqref{NARX pred_sp} and Fig. \ref{ser-par_pic}). A parallel architecture on the other hand can be used to perform multiple step ahead predictions (MSAP) by feeding back the predictions of the OSAP model in a recurrent manner (see equation \eqref{NARX pred_p} and Fig. \ref{par_pic}). For further reading on series-parallel and parallel architectures, the reader is referred to \cite{narendra}.
\begin{equation}\label{NARX pred_sp}
\hat{y}(k+1) = \hat{f}_{NARX}[u(k),..,u(k-n_u+1),y(k),..,y(k-n_y+1)] \\
\end{equation}
%\begin{figure*}[]
%      \centering
%      \includegraphics[scale=0.5]{ser_par_struc.pdf}
%      \caption{Series-Parallel architecture for system identification.}
%      \label{ser-par_pic}
%\end{figure*}
%\begin{figure*}[]
%      \centering
%      \includegraphics[scale=0.5]{par_struc.pdf}
%      \caption{Parallel Architecture for system identification.}
%      \label{par_pic}
%\end{figure*}
\begin{multline}\label{NARX pred_p}
\hat{y}(k+n_{pred}) = \hat{f}_{NARX}[u(k+n_{pred}-1),..,u(k-n_u+n_{pred}), \\
                        \hat{y}(k+n_{pred}-1),..,\hat{y}(k-n_y+n_{pred})].
\end{multline}
The OSAP model is used for training as existing simple training algorithms can be used and once the model becomes accurate for OSAP, it can be converted to a MSAP model recursively in a straightforward manner. The MSAP model can be used for making long term predictions useful for predictive control discussed in section \ref{sim sec}.

\subsection{Extreme Learning Machines}
\begin{figure}[htp]
      \centering
      \includegraphics[scale=0.5]{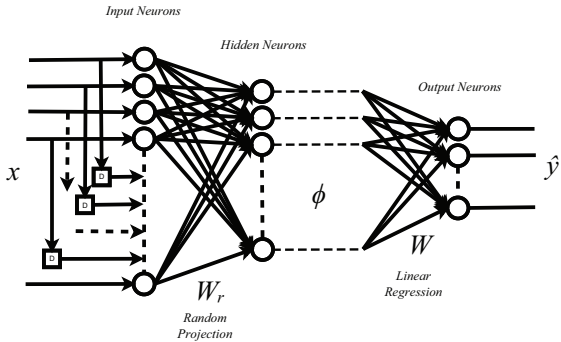}
      \caption{Extreme learning machine model structure.}
      \label{elmfig}
\end{figure}

Extreme Learning Machines is an emerging learning paradigm for multi-class classification and regression problems \cite{4Huang2005,huang12} and has outperformed some state of the art algorithms such as backpropagation neural nets, support vector machines etc. The highlight of ELM is that the training speed is extremely fast with superior generalization capabilities. The key enabler for ELM's training speed is the random assignment of input layer parameters which do not require adaptation to the data. In such a setup, the output layer parameters can be determined analytically using linear least squares (See Fig. \ref{elmfig}). Some of the attractive features of ELM include the universal approximation capability, better generalization, the convex optimization problem of ELM resulting in the smallest training error without getting trapped in local minima, availability of a closed form solution eliminating iterative training \cite{4Huang2005}.

ELM training involves solving the following optimization problem
\begin{equation}\label{ELM_opti}
\min_{W}\left\{\|HW-Y\|^2+\lambda \|W\|^2\right\}
\end{equation}
\begin{equation}\label{}
\phi=H^T=\psi(W_r^Tx(k)+b_r)\in\mathbb{R}^{n_h \times 1}
\end{equation}
where $\lambda$ represents the regularization coefficient, Y represents the vector of outputs, $\psi$ represents the hidden layer activation function (typically sigmoidal or radial basis functions) and $W_r, W$ represents the input and output layer parameters respectively. Here, $n_h$ represents the number of hidden neurons of the ELM model and $H$ represents the hidden layer output matrix.

A prominent feature of ELM is that the nonlinear optimization problem is reduced to a linear parameter estimation problem. This reduction is made possible by the random assignment of the input layer parameters. The matrix $W_r$ consists of randomly assigned elements that maps the input vector to a high dimensional feature space while $b_r\in\mathbb{R}^{n_h}$ is a bias component assigned in a random manner similar to $W_r$. The number of hidden neurons determine the dimension of the transformed feature space. The random parameters can be assigned based on any continuous random distribution \cite{huang12} and remains fixed during the training process. The intuition behind the random parametrization of ELM can be explained as follows. By assigning random weights as described above, along with an activation function, a regressor $\phi$ with several functional combinations (with different order terms) of the input feature vector $x$ is obtained. For a complex problem, more hidden neurons are required which can be thought of as introducing more complex feature mappings. When the dimension of the regressor $\phi$ is sufficiently high with sufficient inclusion of higher order (nonlinear) features, it can be mapped to the outputs linearly. Hence the training reduces to a single step calculation given by equation \eqref{ELM_train}. The ELM decision hypothesis can be expressed as in equation \eqref{elm model_reg}
\begin{equation}\label{ELM_train}
W^*=\left(H^TH + \lambda I \right)^{-1}H^TY.
\end{equation}
\begin{equation}\label{elm model_reg}
f(x)=W^T[\psi(W_r^Tx+b_r)].
\end{equation}

\subsection{Model Development and Validation}
\begin{figure*}[hbtp]
      \centering
      \begin{tabular}{cc}
      \subfloat[A 600 cycle ahead prediction results of the ELM engine model (unseen test data set 1).]{\label{1800_pred_1}\includegraphics[scale=0.61]{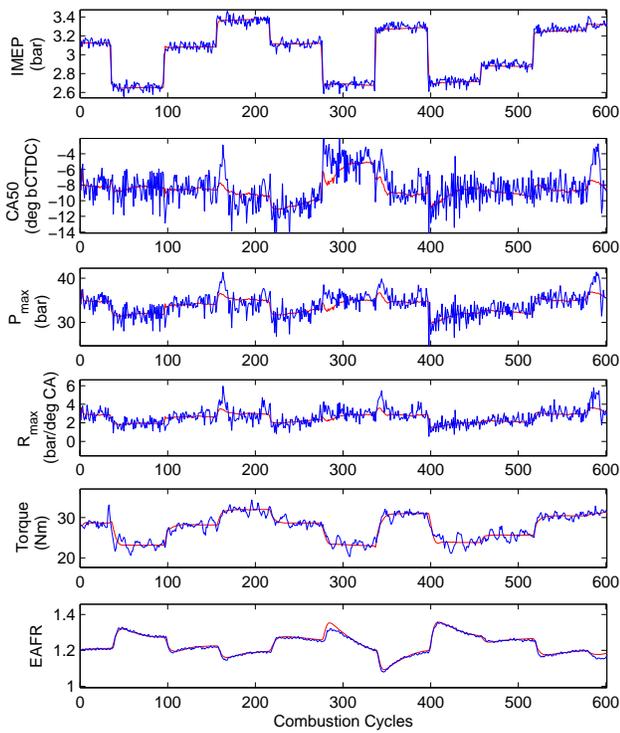}}
      &      \subfloat[A 600 cycle ahead prediction results of the ELM engine model (unseen test data set 2).]{\label{1800_pred_2}\includegraphics[scale=0.61]{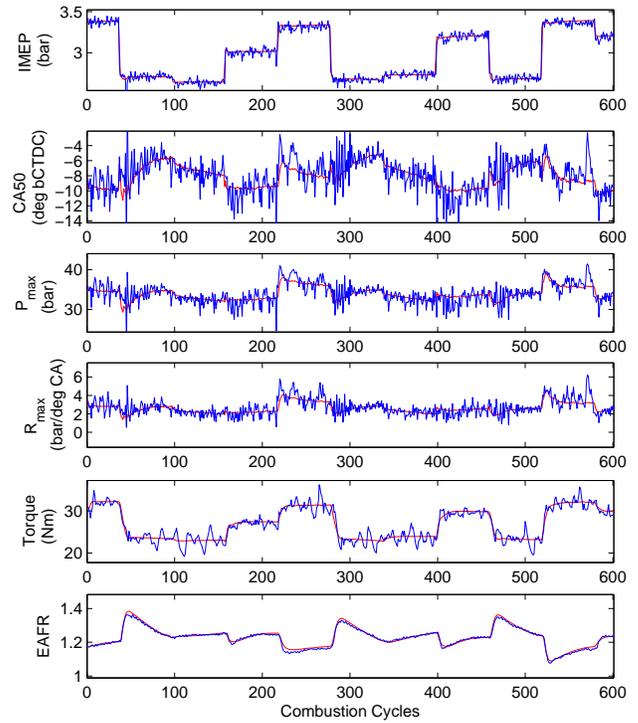}}\\
      \subfloat[A 600 cycle ahead prediction results of the ELM engine model (unseen test data set 3).]{\label{1800_pred_3}\includegraphics[scale=0.61]{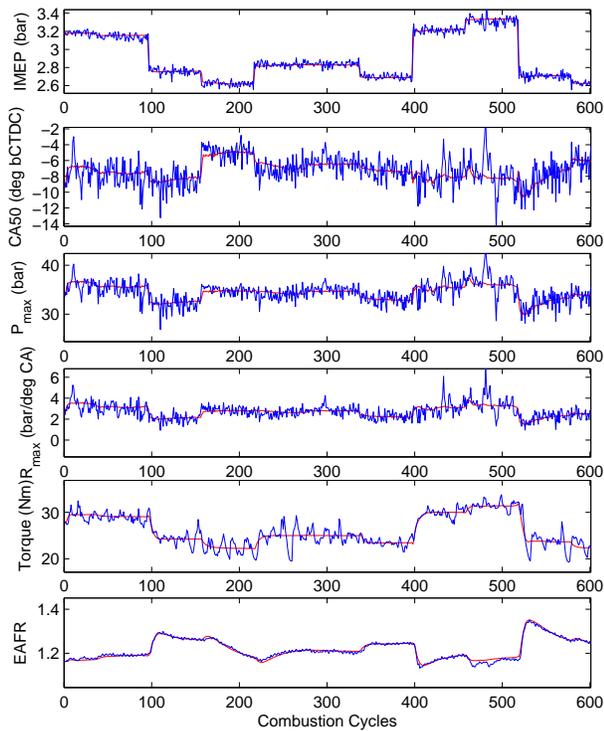}}
      &      \subfloat[A 600 cycle ahead prediction results of the ELM engine model (unseen test data set 4).]{\label{1800_pred_4}\includegraphics[scale=0.61]{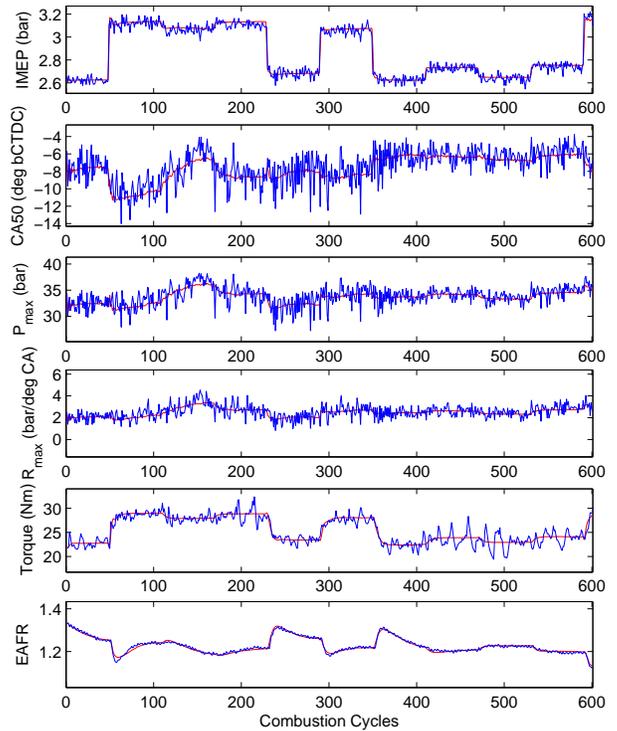}}\\
      \end{tabular}
      \caption{Multi-step prediction performance of the ELM HCCI engine model. The prediction is shown in red while the actual data is shown in blue. The corresponding model inputs (engine control inputs) such as FM, EVC and SOI are not shown as it is common to both prediction and actual data.}
      \label{}
\end{figure*}

The engine experimental data is split into training and testing sets. The training set consists of about 16000 cycles of data while the testing set consists of about 7000 cycles. In order to validate multi-step ahead prediction performance, a separate MSAP-testing set containing about 2400 cycles of data is used. Firstly, the model hyper-parameters are tuned using a cross-validation approach \cite{vijay_asoc,vijay_springer} is employed where a small subset of the training set is used to train the model with several combinations of hyper-parameters and testing its performance on the rest of the unseen training set. After tuning the model hyper-parameters using cross-validation, an optimal model structure is determined. The optimal HCCI engine model consists of 20 hidden neurons with a regularization coefficient ($\lambda$) of 0.001 and a system order ($n_0$) identified to be 1. The model is retrained with the optimal hyper-parameters and evaluated on an unseen test data set. The testing and MSAP-testing data sets are never seen by the models during training.

The models developed in this section are intended for use in a predictive control framework where real time predictions are required in the engine ECU. In order to evaluate the prediction capability of the models, a root mean squared error (RMSE) given by equation \eqref{rmse_eqn} is used. The models are trained using a series-parallel architecture \cite{narendra} and one-step ahead prediction performance is evaluated on the unseen testing data set. Finally, the model predictions are evaluated for $n_{pred}$ steps ahead in time using a separate input sequence. Both the testing set as well as the input sequence for multi-step ahead predictions are never seen by the model during training. This gives a good measure of the long term predictions as well as the generalization capability of the models to unseen situations.
\begin{equation}\label{rmse_eqn}
RMSE=\sqrt{\frac{1}{N}\sum_{i=1}^{N}\sum_{j=1}^{y_d}(y_j^i-\hat{y}_j^i)^2}.
\end{equation}
The test RMSE for one-step ahead prediction is observed to be 0.1085 while for a 600-step ahead prediction, the test RMSE is observed to be 0.1092. The predictions are summarized in Fig. \ref{1800_pred_1} to Fig. \ref{1800_pred_4} for different unseen MSAP-testing data sets. The RMSE values correspond to data normalized between -1 and +1, and give a good indication of how well the model can predict the dynamics of the HCCI engine. The MSAP predictions at several different engine operating conditions as well as the MSAP RMSE values prove that the model can be used for predicting multiple steps ahead in time and can be used as a predictive model for the HCCI Engine in the MPC framework to be discussed in the next section \ref{mpc_sec}.

\section{Predictive Control formulation using Extreme Learning Machines}\label{mpc_sec}
In this section, the MPC problem is formulated based on general predictive control principles \cite{maciej,maciej_nn} for a class of nonlinear systems using ELM models. The formulation will be applied to the HCCI engine problem in the later sections.

Consider a general class of nonlinear discrete time system,
\begin{eqnarray}
  z(k+1) &=& f(z(k),u(k))\nonumber \\
  y(k) &=& g(z(k))\label{system_defn},
\end{eqnarray}
where $z \in \mathbb{R}^n$, $u \in \mathbb{R}^m$, $y \in \mathbb{R}^p$, $k$ the time index, $n, m$ and $p$ represent the number of states, inputs and outputs respectively. $f(.)$ and $g(.)$ are continuously differentiable and globally Lipschitz \cite{lipsc} nonlinear functions modeled offline using ELM. The models can be linearized around any operating point ($z_0,u_0$) using Taylor's series expansion as follows
\begin{eqnarray*}\label{}
z(k+1) &=& f(z^0(k),u^0(k))+A(z(k)-z^0(k)) \\
        && +B(u(k)-u^0(k))+\epsilon_{z,ho}\\
y(k) &=& g(z^0(k))+C(z(k)-z^0(k))+\epsilon_{y,ho}
\end{eqnarray*}
\begin{eqnarray*}
  A &=& \left[\frac{\partial f}{\partial z}(z(k),u(k))\right]_{z^0(k),u^0(k)} \\
  B &=& \left[\frac{\partial f}{\partial u}(z(k),u(k))\right]_{z^0(k),u^0(k)} \\
  C &=& \left[\frac{\partial g}{\partial z}(z(k),u(k))\right]_{z^0(k),u^0(k)}
\end{eqnarray*}
where $A,B$ and $C$ represent the partial derivatives of the ELM models in the Taylor's expansion, $\epsilon_{z,ho}$ and $\epsilon_{y,ho}$ represent higher order terms. The linearized model can be further written as
\begin{eqnarray}
z(k+1) &=& A z(k)+B u(k) + d_1(k) \nonumber\\
y(k)   &=& C z(k) + d_2(k)\label{lin_sys}
\end{eqnarray}
\begin{eqnarray*}
  d_1 &=& f(z^0(k),u^0(k))-(A z^0(k) + B u^0(k)) + \epsilon_{z,ho} \\
  d_2 &=& g(z^0(k),u^0(k))-C z^0(k) + \epsilon_{y,ho}.
\end{eqnarray*}
The following subsections describe the calculation of the system matrices $A,B$ and $C$ followed by the MPC optimization problem formulation.

\subsection{Calculation of System Matrices}
The matrices $A,B$ and $C$ in equation \eqref{lin_sys} can be determined by exploiting the structure of the ELM model as follows. Let the augmented input vector to ELM be given by $x(k)=[u(k),z(k)]^T\in\mathbb{R}^{n+m}$. The matrices $A,B$ and $C$ can be determined by calculating the Jacobian of $f(.)$ and $g(.)$ with respect to the augmented input vector $x(k)$. The ELM model structure can be expressed as
\begin{equation}\label{elm model}
\hat{z}(k+1)=W^T[\psi(W_r^Tx(k)+b_r)]
\end{equation}
where $\psi$ represents the hidden layer activation function and $W_r\in\mathbb{R}^{n+m \times n_h}, W\in\mathbb{R}^{n_h \times n}$ represents the input and output layer parameters respectively. Here, $n_h$ represents the number of hidden neurons of the ELM model, $\phi(k)=\psi(W_r^Tx(k)+b_r)\in\mathbb{R}^{n_h \times 1}$ represents the hidden layer output matrix as termed in literature (see Fig. \ref{elmfig}). As mentioned earlier, the matrix $W_r$ consists of randomly assigned elements that maps the input vector to a high dimensional feature space while $b_r\in\mathbb{R}^{n_h}$ is a bias component assigned in a random manner similar to $W_r$. The elements can be assigned based on any continuous random distribution \cite{huang12} and remains fixed during the learning process. The number of hidden neurons determine the dimension of the transformed feature space and the hidden layer is equipped with a nonlinear activation function similar to traditional neural network architecture. In this paper, a sigmoidal activation function is considered.

Note that the ELM model in \eqref{elm model} is defined for inputs and outputs normalized to lie between $[-1,+1]$. Expressing the model in \eqref{elm model} along with the normalization and de-normalization terms,
\begin{eqnarray}\label{elm model_actual}
\nonumber \hat{z}(k+1) &=& z_{min}+\left(\frac{z_{max}-z_{min}}{2}\right)\left\{1+W^T \phi(k) \right\} \\
\phi(k) &=& \frac{1}{1+e^{-\left\{W_r^T \left[2\left(\frac{x(k)-x_{min}}{x_{max}-x_{min}} \right)-1 \right]+b_r \right\}}}.
\end{eqnarray}
Then, Jacobian matrix $\frac{\partial f}{\partial x}$ can be derived (ignoring the time index $k$) as
\begin{equation}\label{}
\frac{\partial f}{\partial x} = \left(\frac{z_{max}-z_{min}}{2}\right)W^T \frac{\partial \phi}{\partial x} \\
\end{equation}
where
\begin{multline*}\label{}
\frac{\partial\phi_i}{\partial x_j} = \\
\frac{2W_r(j,i)e^{-\left\{W_{r,1}^T \left[2\left(\frac{x-x_{min}}{x_{max}-x_{min}} \right)-1 \right]+b_{r,1} \right\}}}{(x_{max,j}-x_{min,j})\left(1+e^{-\left\{W_{r,1}^T \left[2\left(\frac{x-x_{min}}{x_{max}-x_{min}} \right)-1 \right]+b_{r,1} \right\}}\right)^2},
\end{multline*}
$W_{r,i}$ represents the $i^{th}$ column of $W_r$ and $b_{r,i}$ represents the $i^{th}$ element of $b_r$, $\frac{\partial \phi}{\partial x}\in\mathbb{R}^{n_h \times n+m}$. Since the augmented vector is defined as $x(k)=[u(k),z(k)]^T$, the matrices $A$ and $B$ can be extracted from the Jacobian $\frac{\partial f}{\partial x}$ as
\begin{equation}\label{}
\left[\frac{\partial f}{\partial x} \right]_{n \times n+m} = [B_{n \times m}|A_{n \times n}].
\end{equation}
A similar jacobian calculation can be done for $g(.)$ and also for other type of activation functions $\psi(.)$. The above calculations are algebraic and can be efficiently performed online.

\subsection{MPC Optimization Problem}
The goal of MPC is to force the system output $y(k)$ track a given reference $r(k)$ and also penalize any large excursion in the input signal $u(k)$. This is obtained by solving the following optimization problem at every time instant $k$
\begin{equation}\label{cost}
J(k) = \sum_{j=1}^{N_y} \| r(k+j|k)-y(k+j|k) \|^2_{Q_1} + \sum_{j=0}^{N_u-1} \| \Delta u(k+j|k) \|^2_{Q_2}
\end{equation}
\begin{equation}\label{first_const}
 \text{subjected to}
  \begin{cases}
      u_{min} \leq u(k+j|k) \leq u_{max}\\
      \Delta u_{min} \leq \Delta u(k+j|k) \leq \Delta u_{max}\\
      y_{min} \leq y(k+j|k) \leq y_{max}
  \end{cases}
\end{equation}
where $r(k+j|k), y(k+j|k), \Delta u(k+j|k)$ represents the reference, system output and control increment respectively. The argument $(k+j|k)$ indicates the signal from time index $k$ until $k+j$ being used for solving the optimization problem at time index $k$. Here $\Delta u(k+j|k) = u(k+j|k) - u(k-1+j|k)$, $N_y$ and $N_u$ are prediction horizon ($N_y \geq 1$) and control horizon ($0<N_u \leq N_y$) respectively, $\|.\|_{Q_1}$ and $\|.\|_{Q_2}$ denote weighted Euclidean norm weighted by matrices $Q_1$ and $Q_2$ respectively. The minimum and maximum constraints for $u$, $\Delta u$ and $y$ are given as in equation \eqref{first_const}.

By defining the following vectors,
\begin{eqnarray*}
Y(k) &=& [y(k+1|k),y(k+2|k),..,y(k+N_y|k)]^T \\
\Delta U(k) &=& [\Delta u(k|k),\Delta u(k+2|k),..,\Delta u(k+N_u-1|k)]^T \\
R(k) &=& [r(k+1|k),r(k+2|k),..,r(k+N_y|k)]^T
\end{eqnarray*}
and calculating $y(k+j|k)$ recursively using equation \eqref{lin_sys}, the vector $Y(k)$ can be expressed as
\begin{multline}\label{}
Y(k) = \mathcal{Z}(k)z(k) + \mathcal{U}(k) \Delta U(k) + \mathcal{V}(k) u(k-1) \\
+ \mathcal{D}_1(k) d_1(k) + \mathcal{D}_2(k) d_2(k)
\end{multline}
where \begin{small}
\[ \mathcal{U} = \left [ \begin{array}{cccc}
CB &. &. & 0\\
CB+CAB &. &. & .\\
. &. &. & .\\
CB+..+CA^{N_u-1}B &. &. & CB\\
CB+..+CA^{N_u}B &. &. & CB+CAB\\
. &. &. & .\\
CB+..+CA^{N_y-1}B &. &. & CB+..+CA^{N_y-N_u}B\\
\end{array} \right ] \]
\begin{center}
$\mathcal{V} = \left [ \begin{array}{c}
CB \\
CB+CAB \\
. \\
.\\
CB+CA^{N_y-1}B
\end{array} \right ],
\mathcal{Z} = \left [ \begin{array}{c}
CA \\
CA^2 \\
. \\
.\\
CA^{N_y}
\end{array} \right ]$

$\mathcal{D}_1 = \left [ \begin{array}{c}
C \\
C+CA \\
. \\
.\\
C+CA^{N_y-1}
\end{array} \right ],
\mathcal{D}_2 = \left [ \begin{array}{c}
I_p\\
I_p\\
.\\
.\\
I_p
\end{array} \right ]$.
\end{center}
\end{small}

$\\
Q_1\in \mathbb{R}^{N_y p \times N_y p}, Q_2\in\mathbb{R}^{N_u m \times N_u m}, Y\in\mathbb{R}^{N_y p \times 1}, \\
\Delta U\in\mathbb{R}^{N_u m \times 1}, R\in\mathbb{R}^{N_y p \times 1}, \mathcal{U}\in\mathbb{R}^{N_u p \times N_u m}, \\
\mathcal{V}\in\mathbb{R}^{N_y p \times m}, \mathcal{Z}\in\mathbb{R}^{N_y p \times n},\mathcal{D}_1\in\mathbb{R}^{N_y p \times n}, \mathcal{D}_2\in\mathbb{R}^{N_y p \times p}$.

The optimization problem in equation \eqref{cost} can now be expressed in vector form as
\begin{equation}\label{redefined_cost}
\min_{\Delta U(k)} \left \{ \Delta Y^T(k) Q_1 \Delta Y(k) + \Delta U^T(k) Q_2 \Delta U(k) \right \}
\end{equation}
where
\begin{small}
\begin{eqnarray}
\nonumber \Delta Y(k) &=& R(k)-Y(k) \\
\nonumber            &=& R(k)-\mathcal{Z}(k)z(k) - \mathcal{U}(k) \Delta U(k) \\
\nonumber            &&  - \mathcal{V}(k) u(k-1) - \mathcal{D}_1(k) d_1(k) \\
           &&  - \mathcal{D}_2(k) d_2(k).
\end{eqnarray}
\end{small}
Expressing as a quadratic programming problem, the above formulation reduces to
\begin{equation}\label{qp_cost}
\min_{\Delta U(k)} \left \{ \frac{1}{2} \Delta U^T(k) W_1(k) \Delta U(k) + W_2^T(k) \Delta U(k) \right \}
\end{equation}
\begin{equation}\label{qp_const}
 \text{subjected to   }
      E(k) \leq F(k)
\end{equation}
where
\begin{small}
\begin{eqnarray}
\nonumber W_1(k) &=& 2[\mathcal{U}^T(k) Q_1 \mathcal{U}(k) + Q_2]\\
\nonumber W_2(k) &=& -2\mathcal{U}^T(k) Q_1 [R(k)-\mathcal{Z}(k)z(k) \\
\nonumber        && - \mathcal{V}(k) u(k-1) - \mathcal{D}_1(k) d_1(k) - \mathcal{D}_2(k) d_2(k)]\\
\nonumber E(k) &=& [I_{N_um} \text{ }-I_{N_um} \quad H \text{ } -H \quad \mathcal{U}(k) \text{ } -\mathcal{U}(k)]^T \\
\nonumber F(k) &=& \left[\begin{array}{c} \Delta U_{max} \\ -\Delta U_{min} \\ U_{max}-U_{k-1} \\ -\{U_{min}-U_{k-1}\} \\ \\
Y_{max}-\mathcal{Z}(k)z(k) - \mathcal{V}(k) u(k-1) \\ - \mathcal{D}_1(k) d_1(k) - \mathcal{D}_2(k) d_2(k) \\
\\
-\{Y_{min}-\mathcal{Z}(k)z(k) - \mathcal{V}(k) u(k-1) \\ - \mathcal{D}_1(k) d_1(k) - \mathcal{D}_2(k) d_2(k)\} \\
\end{array} \right] \\
\nonumber H &=& \left[\begin{array}{ccccc}
I_m & 0 & . & . & 0 \\
I_m & I_m & . & . & 0 \\
. & . & . & . & . \\
. & . & . & . & . \\
I_m & I_m & . & . & I_m \\
\end{array}\right]
\end{eqnarray}
\end{small}
\begin{eqnarray*}
U_{min} &=& [u_{min} \quad u_{min}..u_{min}]^T \\
U_{max} &=& [u_{max} \quad u_{max}..u_{max}]^T \\
\Delta U_{min} &=& [\Delta u_{min} \quad \Delta u_{min}..\Delta u_{min}]^T \\
\Delta U_{max} &=& [\Delta u_{max} \quad \Delta u_{max}..\Delta u_{max}]^T \\
U_{k-1} &=& [u(k-1) \quad u(k-1)..u(k-1)]^T \\
Y_{min} &=& [y_{min} \quad y_{min}..y_{min}]^T \\
Y_{max} &=& [y_{max} \quad y_{max}..y_{max}]^T.
\end{eqnarray*}
$I_m$, $I_p$ and $I_{N_um}$ represents identity matrices in $\mathbb{R}^{m \times m}$, $\mathbb{R}^{p \times p}$ and $\mathbb{R}^{N_um \times N_um}$ respectively. The above problem can be solved using efficient quadratic programming solvers \cite{maciej}. The first value of the optimal control increment $\Delta U(k)$ is applied at the present time instant $k$ and the optimization is solved again for the next time index. This is done to handle any model inaccuracies and external disturbances.

\section{Application to Control of a Gasoline Homogeneous Charge Compression Ignition Engine}\label{sim sec}
The goal of this section is to design a real-time control algorithm that can control the HCCI engine to track a given reference command in IMEP and CA50. The IMEP reference represents the driver's load demand while CA50 reference is calculated a priori for achieving a sweet spot between high efficiency, low emissions and stability in combustion. The procedure for calculating reference IMEP and CA50 is outside the scope of this article and the reader is referred to \cite{set_point1,set_point2}. For the controller analysis considered in this paper, it is assumed that the reference commands are available and the goal is to determine the optimal control trajectories of FM, EVC and SOI that tracks the reference commands.

\begin{figure*}[htp]
      \centering
      \includegraphics[scale=0.6]{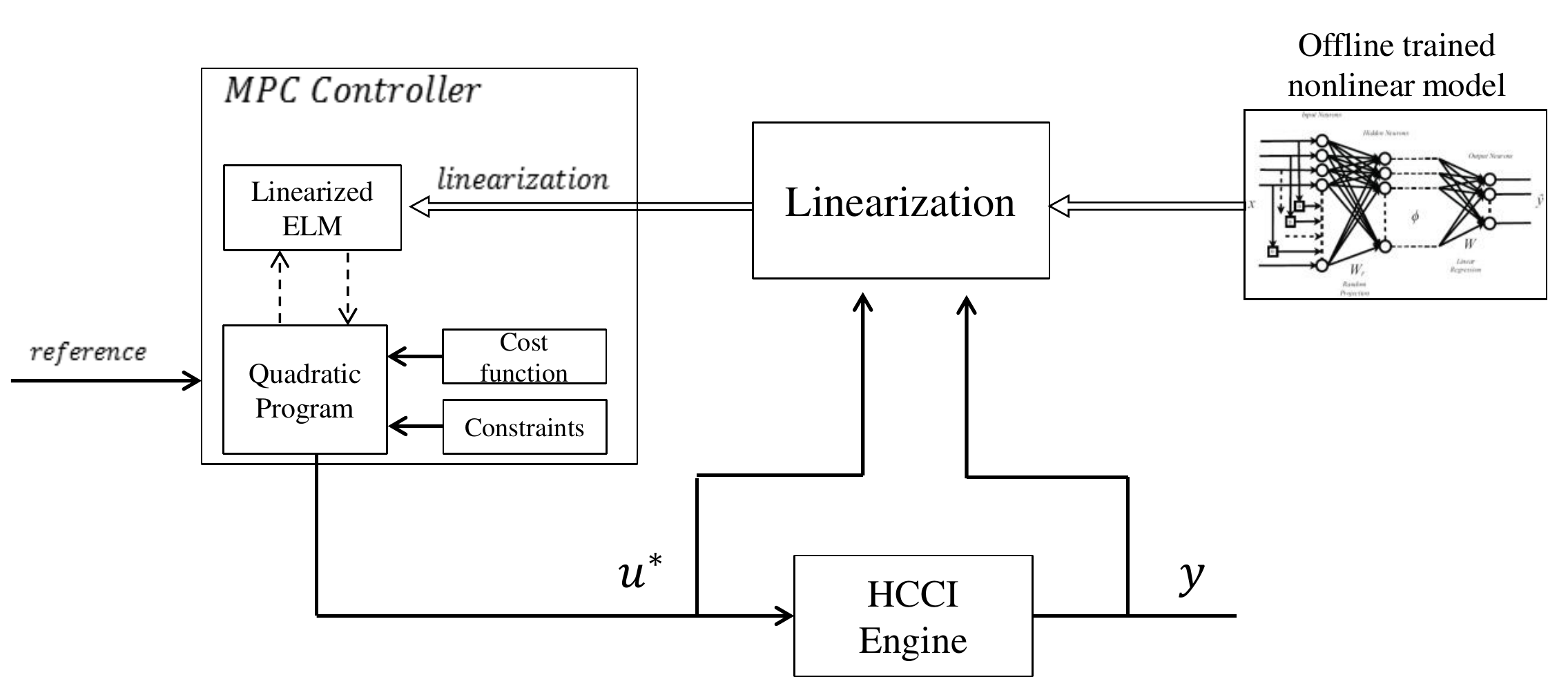}
      \caption{A model predictive control framework showing the use of offline trianed nonlinear engine model for reference tracking.}
      \label{MPC_framework}
\end{figure*}

The MPC framework is shown in Fig. \ref{MPC_framework}. An offline trained nonlinear model of the HCCI engine, developed using ELM in section \ref{modeling_sec}, is used as a proxy for the HCCI engine. At every operating point, the model is linearized and the system matrices are used to formulate a quadratic program as described in section \ref{mpc_sec}. The engine hardware limitations as well as operating constraints are included in the optimization problem. The reference command includes optimal set points of IMEP and CA50 that is to be tracked by the engine. At every time instant, the solver receives the present state information ($z(k)$) from sensor measurements and attempts several control trajectories of FM, EVC and SOI. The optimal trajectory that minimizes the tracking error and control excursion is given to the engine as input.

As mentioned previously, the control knobs for the HCCI engine such as FM, EVC and SOI vary the amount of EGR trapped in the cylinder. The EGR influences the temperature and concentration of the mixture, affecting IMEP and CA50 of the combustion event. The physical states of the system can include a combination of temperature, concentration, pressure of the gas mixture before combustion but a direct measurement of in-cylinder states is not feasible for production engines. In-cylinder pressure is the only available measurement from which input-output models were built in section \ref{modeling_sec}. As a consequence, for the control problem considered here, the states, inputs and outputs of the system are given by
\begin{eqnarray*}
z(k) &=& [IMEP(k) \quad CA50(k) \quad P_{max}(k) \\
     &&  \quad R_{max}(k) \quad T_b(k) \quad \lambda(k)]^T \in\mathbb{R}^6 \\
u(k) &=& [FM(k) \quad EVC(k) \quad SOI(k)]^T \in\mathbb{R}^3 \\
y(k) &=& [IMEP(k) \quad CA50(k)]^T \in\mathbb{R}^2.
\end{eqnarray*}
The dynamic equations of the model can be expressed as in \eqref{system_defn}. This system represents a nonlinear multi-input multi-output model structure which makes control design very challenging. Further, the engine exhibits both system level and operation level constraints which makes traditional linear control methods such as PID, LQ etc. ineffective. Further, additional challenges inherent to the HCCI engine include high noise amplitude levels, high sensitivity to disturbances, fast operation and tight tracking of the reference CA50. As mentioned earlier, an MPC based approach has sufficient leeway to handle the above challenges.

MPC technology is well developed for linear systems with computationally efficient solvers. However, application of MPC to a nonlinear system such as the HCCI engine involves solving a quadratic programming subproblem iteratively at every time instant. To give an idea of the engine sampling period, a quadratic program problem needs to be solved every 48 milliseconds when the engine operates at 2500 RPM. The non-availability of efficient solvers for sequential quadratic programming problem in real-time and the inability to guarantee a global optimal solution for such problems constitute the major bottleneck for nonlinear MPC techniques. However, if the system is not highly nonlinear, a linearized model can be used to obtain suboptimal control laws in real-time using quadratic programming. A quadratic programming is relatively manageable than solving it sequentially for a constrained nonlinear programming problem and hence more suitable for online applications. However, for the considered high-speed engine application, there is very little time even for a simple but generic quadratic programming solver. In the next section, a fast quadratic programming approach is described that makes use of the problem's strictly convex objective function to speed up convergence.

\subsection{Fast Quadratic Programming}
The MPC problem involves solving a quadratic programming subproblem of the form given by equation \eqref{qp_cost}. The generic quadratic programming solver demands computation and memory owing to iterative interior point, active sets or trust region approximations and becomes infeasible for implementation on the engine ECU. In order to reduce the computation and memory, an efficient and fast QP algorithm is adopted \cite{qp_nn}. The algorithm can be derived as follows.

Restating the MPC quadratic programming problem from \eqref{qp_cost},
\begin{equation}\label{}
\min_{\Delta U(k)} \left \{ \frac{1}{2} \Delta U^T(k) W_1(k) \Delta U(k) + W_2^T(k) \Delta U(k) \right \}
\end{equation}
\begin{equation}\label{}
 \text{subjected to   }
      E(k) \Delta U(k) - F(k) \leq 0
\end{equation}
where $W_1(k)$ is a symmetric positive definite matrix and the objective function is strictly convex. The lagrangian dual problem can be formulated as
\begin{multline}\label{}
\max_{\lambda_L} \inf \bigg \{ \frac{1}{2} \Delta U^T(k) W_1(k) \Delta U(k) + W_2^T(k) \Delta U(k) \\
+ \lambda_L(E(k) \Delta U(k) - F(k))\bigg \}
\end{multline}
where $\lambda_L \geq 0$. Taking the derivative with respect to $\Delta U$,
\begin{equation}\label{dual_opt1}
W_1(k) \Delta U(k) + W_2(k) + E^T(k) \lambda_L = 0.
\end{equation}
The dual problem can be expressed as
\begin{multline}\label{dual_MPC}
\max_{\lambda_L} \bigg \{ \frac{1}{2} \Delta U^T(k) W_1(k) \Delta U(k) + W_2^T(k) \Delta U(k) \\
+ \lambda_L(E(k) \Delta U(k) - F(k)) \bigg \}
\end{multline}
subjected to
\begin{equation*}\label{}
W_1(k) \Delta U(k) + W_2(k) + E^T(k) \lambda_L = 0
\end{equation*}
\begin{equation*}
      y \geq  0.
\end{equation*}
Since $W_1(k)$ is positive definite, $W_1(k)^{-1}$ exists and equation \eqref{dual_opt1} can be solved as follows
\begin{equation}\label{opt_sol1}
\Delta U^*=-W_1(k)^{-1}(W_2(k) + E^T(k) \lambda_L).
\end{equation}
Following a direct substitution of \eqref{opt_sol1} in \eqref{dual_MPC}, the problem becomes
\begin{equation}\label{fast_qp}
\max_{\lambda_L} \left \{ \frac{1}{2} \lambda_L^T \Lambda_1 \lambda_L + \lambda_L^T \Lambda_2 - \frac{1}{2} W_2^T(k) W_1^{-1}(k) W_2(k) \right \}
\end{equation}
\begin{equation}\label{fast_qp_con}
 \text{subjected to   }
      y \geq 0
\end{equation}
where $\Lambda_1=-E(k) W_1^{-1}(k) E^T(k)$ and $\Lambda_2 = - F(k) - E(k) W_1^{-1}(k) W_2(k)$. The problem in \eqref{fast_qp} can be solved using a simple gradient ascent method as follows.
\begin{equation}\label{}
\lambda_{L_{k+1}}=\lambda_{L_k}+\lambda_{step}(\Lambda_1 \lambda_L + \Lambda_2).
\end{equation}
In order to satisfy the constraint in \eqref{fast_qp_con}, the following modification is made
\begin{equation}\label{qp_fast_final}
\lambda_{L_{k+1}}=max(\lambda_{L_k}+\lambda_{step}(\Lambda_1 \lambda_L + \Lambda_2),0)
\end{equation}
where $\lambda_{step}$ defines the step size. The solution of \eqref{qp_fast_final} gives the optimal $\lambda_L^*$ which can be substituted in \eqref{opt_sol1} to get the optimal MPC output $\Delta U^*$. The strictly convex property of the MPC problem is made use of in solving the quadratic programming subproblem efficiently.

\subsection{Simulation Setup}
For the study considered here, the offline trained nonlinear model is considered as the true HCCI engine plant for which the MPC controller is designed. The goal of the controller is to track the reference IMEP and CA50 trajectories with minimum error and with minimum change in control input (see \eqref{cost}). The reference trajectory is designed offline depending on the allowable operating regions of HCCI and the valid region of the HCCI engine model. The step references are designed to vary between 2.6 and 3.2 bar IMEP and between -4 and -10 deg CA50 defined in degrees before combustion TDC. This covers most of the naturally aspirating HCCI operating range at 1800 RPM. The MPC constraints are defined as follows. It should be noted that the objective function and constraints can be modified in a straightforward manner to include more performance metrics such as emissions, combustion roughness, knock levels etc.
\begin{eqnarray*}
u_{min}&=&[19 \quad -121 \quad 272]^T \\
u_{max}&=&[25 \quad -100 \quad 375]^T \\
\Delta u_{min}&=&[-6 \quad -22 \quad -103]^T \\
\Delta u_{max}&=&[6 \quad 22 \quad 103]^T \\
y_{min}&=&[2.1 \quad -14]^T \\
y_{max}&=&[3.55 \quad -2]^T.
\end{eqnarray*}

\subsection{High Amplitude Noise of HCCI}
\begin{figure*}[]
      \centering
      \includegraphics[scale=0.9]{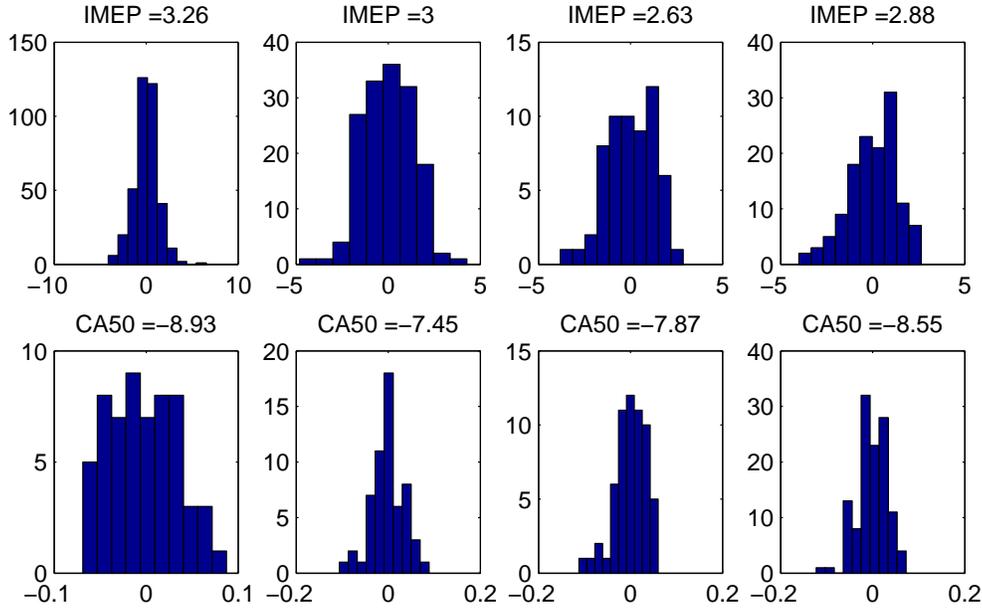}
      \caption{Noise characteristics of CA50 (top rows) and IMEP (bottom rows) as observed in the HCCI engine experimental data for different operating conditions.}
      \label{noise}
\end{figure*}

As mentioned earlier, combustion phasing is an important quantity that indicates engine performance. Combustion is not an instantaneous process but extends over a period of time. Significant research has been done in determining an appropriate parameter that summarize the combustion event and that has good indication over phasing \cite{ca50_1,ca50_2}. It was widely accepted that CA50 was a robust indicator for combustion phasing. However, CA50 is characterized by a highly random process involving noise in the in-cylinder pressure measurements, variability in gas mixing, heat transfer, difference in cylinder-to-cylinder design and cycle-to-cycle variations \cite{thring,misf1,ca50_1}. As a consequence, the variability in CA50 is high. The IMEP is a more averaged effect of the pressure characteristics in the cylinder and thus its variability is lesser. The noise distribution of IMEP and CA50 at certain representative operating conditions are summarized in Fig. \ref{noise}. It can be seen that the noise is almost Gaussian with zero mean. The top rows represent noise in CA50 signals at different IMEP conditions while the bottom rows represent noise in IMEP measurements at different CA50 conditions. The MPC framework considered in this work involves linearization of the models at every operating point defined by sensor measurements. In order to simulate the controller's robustness to the high amplitude noise seen in HCCI engine measurements, white noise signals of zero mean and variances 0.0012 and 1.76 are injected at the outputs of the IMEP and CA50 model respectively and the models are linearized around the noisy measurements experienced in real engine operation.

\subsection{Results And Discussion}
\begin{figure*}[hbtp]
      \centering
      \begin{tabular}{cc}
      \subfloat[State trajectories of the HCCI engine model (with noise) using MPC control (case 1: constraint on $R_{max}$ not enforced in the MPC formulation).]{\label{y1}\includegraphics[scale=0.6]{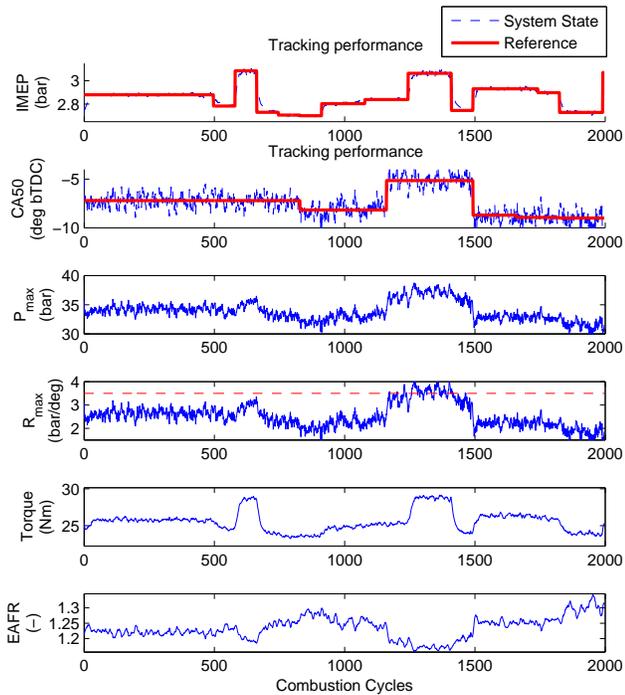}}
      &      \subfloat[Control trajectories of the HCCI engine model (with noise) using MPC control (case 1 constraint on $R_{max}$ not enforced in the MPC formulation). The upper and lower limits of each actuator is shown in dotted red.]{\label{u1}\includegraphics[scale=0.6]{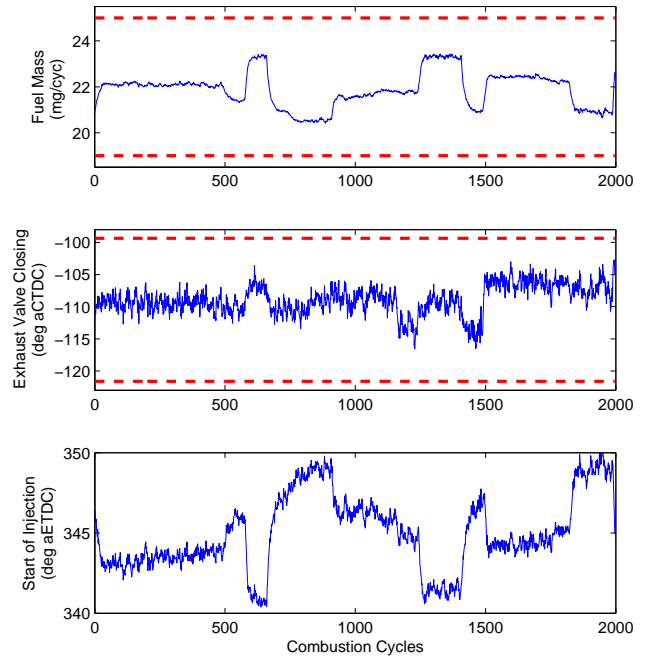}}\\
      \subfloat[State trajectories of the HCCI engine model (with noise) using MPC control (case 2 constraint on $R_{max}$ enforced in the MPC formulation).]{\label{y2}\includegraphics[scale=0.6]{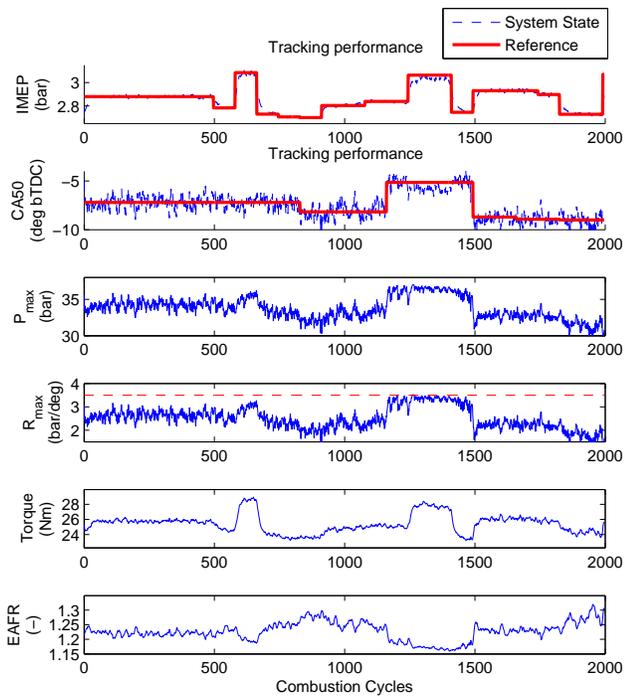}}
      &      \subfloat[Control trajectories of the HCCI engine model (with noise) using MPC control (case 2 constraint on $R_{max}$ enforced in the MPC formulation). The upper and lower limits of each actuator is shown in dotted red.]{\label{u2}\includegraphics[scale=0.6]{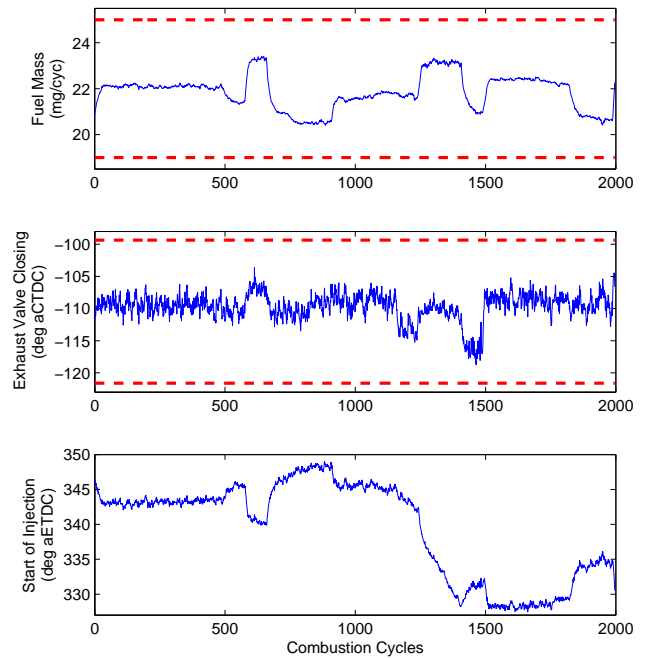}}\\
      \end{tabular}
      \caption{Evaluation of MPC controller using HCCI engine simulator.}
      \label{}
\end{figure*}

In this section, the results of the ELM based MPC results are summarized. A simulation is conducted with a Matlab real-time target bypassing communication with the engine ECU. The ELM model is observed to compactly represent the HCCI engine with about 320 parameters ($W_r\in\mathbb{R}^{n+m \times n_h}, b_r\in\mathbb{R}^{n_h}, W\in\mathbb{R}^{n_h \times n}$). This compactness in comparison to earlier neural network and SVR models is significant \cite{vijay_asoc,vijay_springer} for the MPC application. Further, the fast quadratic programming approach enabled online implementation of the MPC. At every simulation step, the inputs and outputs of the engine model are available as sensor measurements to the MPC system. The HCCI model is linearized around sampled states and is used in the online MPC algorithm to determine the optimal control increment ($\Delta U^*$), i.e., the optimal fuel mass, EVC and SOI to be given to the engine to track the reference commands. After performing several trial and error simulations, the prediction and control horizons take values $N_y=3$ and $N_u=3$, tuned for minimum tracking error. Similarly, the gain matrices are tuned to be
\begin{center}
$Q_1 = \left [ \begin{array}{cccccc}
500 & 0 & 0 & 0 & 0 & 0 \\
0 & 1 & 0 & 0 & 0 & 0 \\
0 & 0 & 500 & 0 & 0 & 0 \\
0 & 0 & 0 & 1 & 0 & 0 \\
0 & 0 & 0 & 0 & 500 & 0 \\
0 & 0 & 0 & 0 & 0 & 1
\end{array} \right ]$
\end{center}
\begin{center}
$Q_2 = \left [ \begin{array}{ccccccccc}
20 & 0 & 0 & 0 & 0 & 0 & 0 & 0 & 0\\
0 & 1 & 0 & 0 & 0 & 0 & 0 & 0 & 0\\
0 & 0 & 1 & 0 & 0 & 0 & 0 & 0 & 0\\
0 & 0 & 0 & 20 & 0 & 0 & 0 & 0 & 0\\
0 & 0 & 0 & 0 & 1 & 0 & 0 & 0 & 0\\
0 & 0 & 0 & 0 & 0 & 1 & 0 & 0 & 0\\
0 & 0 & 0 & 0 & 0 & 0 & 20 & 0 & 0\\
0 & 0 & 0 & 0 & 0 & 0 & 0 & 1 & 0\\
0 & 0 & 0 & 0 & 0 & 0 & 0 & 0 & 1
\end{array} \right ].$
\end{center}
The odd diagonal values of $Q_1$ correspond to reducing IMEP tracking error while the even diagonal values correspond to reducing CA50 tracking error. The relative difference in values between IMEP and CA50 indicates the relatively high noise level of CA50 and a smaller weight trying to track the noisy CA50 less aggressively. $Q_2$ is tuned to avoid saturation of the input signals (taken as constraints in the MPC formulation). The gain corresponding to FM is high indicating a large excursion in FM is penalized so that fuel mass is slowly varied. This is done to ensure that both FM and SOI from reaching their actuator limits.

To evaluate the MPC performance, a reference command involving a series of random steps of IMEP and CA50 is considered. The step input to the controller can be thought of as shifting the engine from one set point to another and gives a good idea of the control performance in terms of response speed, transient and steady state performances. It can be observed from Fig. \ref{y1} that the controller is able to track the considered reference. The time taken by the controller to switch the engine from one set point to the next is about 5 cycles for a small step to about 20 cycles for a large step. By tuning the MPC gains further, the transients of the controller can be shaped to tradeoff between quickness in response to overshoot and minimum oscillations around the reference. Care must be taken while tuning the gains so that the controller does not saturate the actuators or violate the constraints. Also from Fig. \ref{y1}, the transients of the other states of the engine including $P_{max}$, $R_{max}$, engine brake torque and fuel richness in the mixture can be observed. These quantities are obtained from the HCCI engine model for the optimal MPC control trajectories. The control inputs (optimal trajectories) of MPC such as FM, EVC and SOI can be seen in Fig. \ref{u1}. It can be observed that none of the actuators violate the specified limits.

In the above simulation, the constraints are specified only for the actuators and output variables such as IMEP and CA50. However in real situations, typically the engine operation is constrained between stability limits such as misfire, knock and ringing. In order to analyze MPC performance with state constraints and stability limits, an additional constraint is defined on $R_{max}$. In Fig. \ref{y1}, the maximum rate of pressure rise ($R_{max}$) exceeds 3.5 bar/deg CA(shown by red dotted line). By including this as a state constraint in the MPC formulation, the second simulation is performed. The MPC performance is summarized in Fig. \ref{y2} and Fig. \ref{u2}. It can be observed that the MPC control is modified so that the engine $R_{max}$ is less than 3.5 bar/deg CA. A direct comparison of the MPC control trajectories in Fig. \ref{u1} and Fig. \ref{u2} shows that when the constraint on $R_{max}$ is active, MPC reduces the SOI command to operate the engine with less pressure rise rates for the given IMEP and CA50 reference commands. This summarizes the capability of the proposed approach to compute a solution for the multi-input multi-output control problem in an optimal manner. It has to be noted that, as the constraint is active, it becomes difficult for MPC to achieve close tracking as the available freedom is restricted by the active constraint. By tuning the gains further, a better tracking may be achieved.

\begin{figure*}[]
      \centering
      \begin{tabular}{cc}
      \subfloat[State trajectories of the HCCI engine model (with noise) using MPC control (case 3: sinusoidal reference commands).]{\label{y3}\includegraphics[scale=0.6]{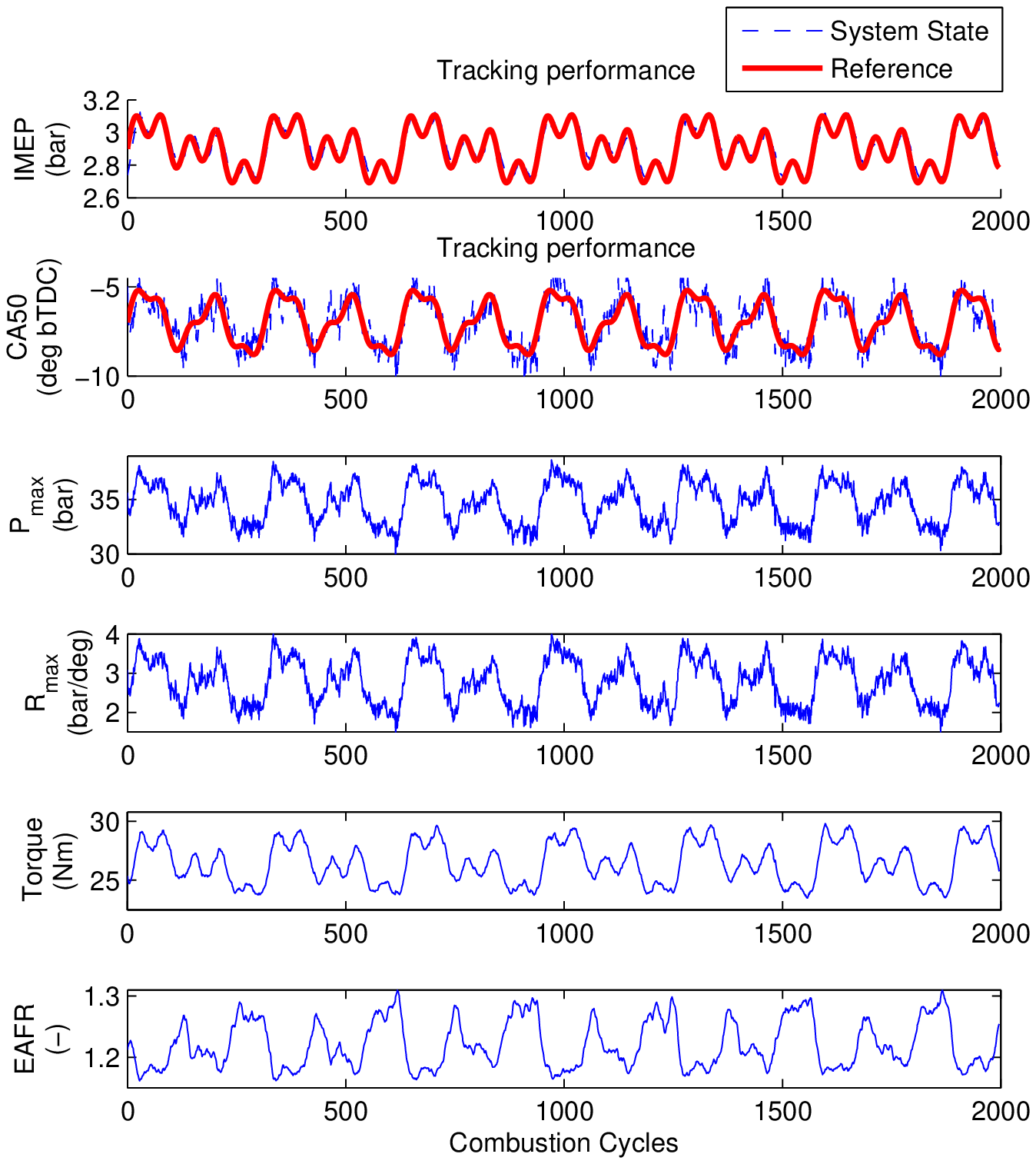}}
      &      \subfloat[Control trajectories of the HCCI engine model (with noise) using MPC control (case 3: sinusoidal reference commands). The upper and lower limits of each actuator is shown in dotted red.]{\label{u3}\includegraphics[scale=0.6]{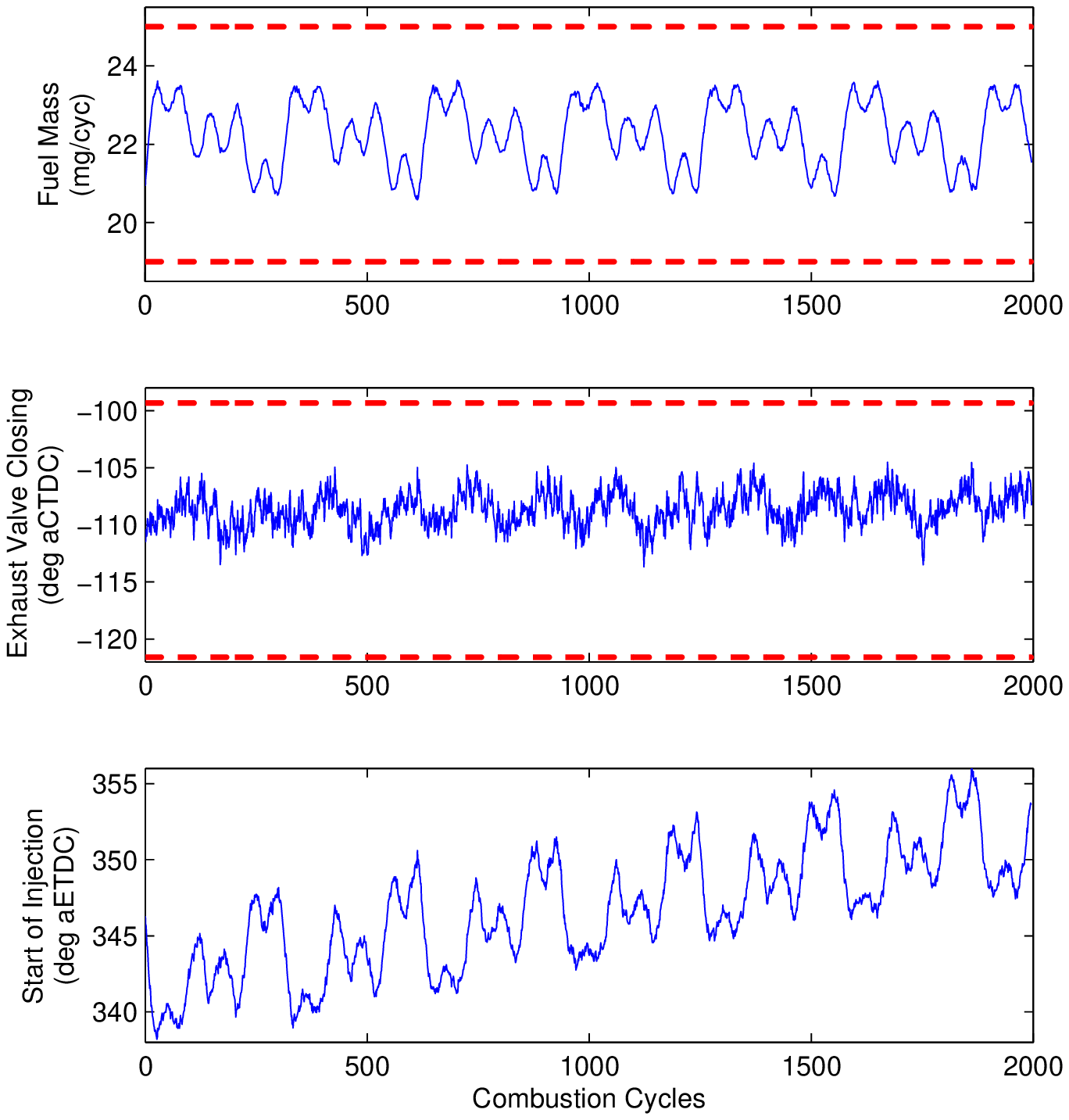}}\\
      \end{tabular}
      \caption{Evaluation of MPC controller using HCCI engine simulator.}
      \label{}
\end{figure*}
Finally, in order to simulate the engine demands in a vehicle application, a slowly varying sinusoidal reference command is given both for IMEP and CA50. This is to simulate the engine power demand while accelerating and slowing down repeatedly. The control performance is summarized in Fig. \ref{y3} and Fig. \ref{u3}. It can be observed that the MPC tracks the given reference to a good degree of accuracy with constraint satisfaction. Hence, the above simulation cases demonstrate the working of ELM based MPC, its manageable computational complexity and its overall potential for real time HCCI engine control.

%\begin{figure}[]
%      \centering
%      \includegraphics[scale=0.64]{y3.eps}
%      \caption{State trajectories of the HCCI engine model (with noise) using MPC control (case 3: sinusoidal reference commands).}
%      \label{y3}
%\end{figure}
%
%\begin{figure}[]
%      \centering
%      \includegraphics[scale=0.64]{u3.eps}
%      \caption{Control trajectories of the HCCI engine model (with noise) using MPC control (case 3: sinusoidal reference commands). The upper and lower limits of each actuator is shown in dotted red.}
%      \label{u3}
%\end{figure}

\section{Conclusions and Future Work}\label{con section}
HCCI engine control is a nonlinear, multi-input multi-output problem with state and actuator constraints which calls for advanced control designs. In this paper, a model predictive control approach has been demonstrated that tracks multiple reference quantities along with constraints on HCCI states and control inputs.

A nonlinear system identification is performed using extreme learning machines and HCCI engine models are developed using experimental data. ELM models are shown to compactly represent the nonlinear dynamics of HCCI and can be used for predicting several steps ahead of time for optimization purposes. An analytical derivative calculation using the structure of ELM models is used which can avoid issues associated with numerical derivatives. The MPC optimization is formulated as a convex problem for which a fast quadratic programming method is used. Simulation studies have been conducted to demonstrate the working of ELM based MPC for the considered HCCI engine problem.

In summary, this article demonstrates a computationally feasible approach to perform MPC for HCCI engines using a nonlinear model in real-time. Future work would focus on extending the method to develop controller for the experimental engine and reformulating MPC to include vehicle level objectives such as fuel economy and emissions.

\section*{Acknowledgment}
This material is based upon work supported by the Department of Energy and performed as a part of the ACCESS project consortium (Robert Bosch LLC, AVL Inc., Emitec Inc.) under the direction of PI Hakan Yilmaz, Robert Bosch, LLC. X. Nguyen is supported in part by NSF Grants CCF-1115769 and ACI-1047871.

\section*{Disclaimer}
This report was prepared as an account of work sponsored by an agency of the United States Government.  Neither the United States Government nor any agency thereof, nor any of their employees, makes any warranty, express or implied, or assumes any legal liability or responsibility for the accuracy, completeness, or usefulness of any information, apparatus, product, or process disclosed, or represents that its use would not infringe privately owned rights.  Reference herein to any specific commercial product, process, or service by trade name, trademark, manufacturer, or otherwise does not necessarily constitute or imply its endorsement, recommendation, or favoring by the United States Government or any agency thereof.  The views and opinions of authors expressed herein do not necessarily state or reflect those of the United States Government or any agency thereof.

%\section*{Acknowledgment}
%This material\footnote{\scriptsize Disclaimer: This report was prepared as an account of work sponsored by an agency of the United States Government.  Neither the United States Government nor any agency thereof, nor any of their employees, makes any warranty, express or implied, or assumes any legal liability or responsibility for the accuracy, completeness, or usefulness of any information, apparatus, product, or process disclosed, or represents that its use would not infringe privately owned rights.  Reference herein to any specific commercial product, process, or service by trade name, trademark, manufacturer, or otherwise does not necessarily constitute or imply its endorsement, recommendation, or favoring by the United States Government or any agency thereof.  The views and opinions of authors expressed herein do not necessarily state or reflect those of the United States Government or any agency thereof.} is based upon work supported by the Department of Energy and performed as a part of the ACCESS project consortium (Robert Bosch LLC, AVL Inc., Emitec Inc.) under the direction of PI Hakan Yilmaz, Robert Bosch, LLC.

\ifCLASSOPTIONcaptionsoff
  \newpage
\fi

\bibliographystyle{IEEEtran}
\bibliography{ELM_MPC}
\newpage
\begin{IEEEbiography}
[{\includegraphics[scale=0.13]{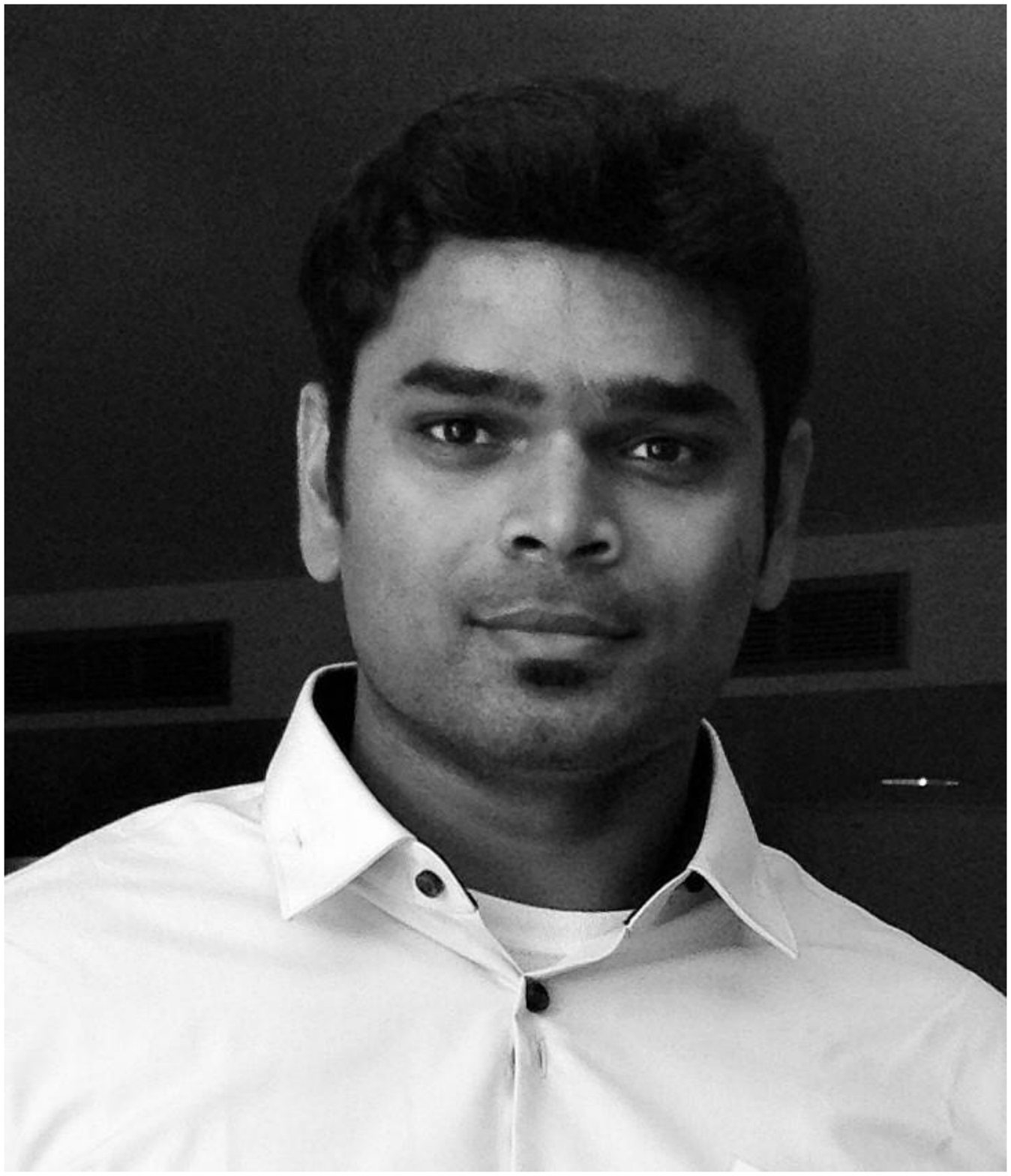}}]{Vijay Manikandan Janakiraman} received the bachelor's degree in Mechanical Engineering (2007) from Sri Venkateswara College of Engineering, Chennai, India. He received Master's degrees in Mechanical Engineering (2008) and Electrical Engineering - Systems (2013) and the Ph.D. degree in Mechanical Engineering (2013) from the University of Michigan at Ann Arbor, MI, USA.

Since August 2013, he has been working as a research scientist with the Data Sciences Group, Intelligent Systems Division at the NASA Ames Research Center, Moffett Field, CA, USA. His current research interests include machine learning, data mining in high dimensional time series, dynamical systems, optimization and control.
\end{IEEEbiography}
\begin{IEEEbiography}
[{\includegraphics[scale=0.13]{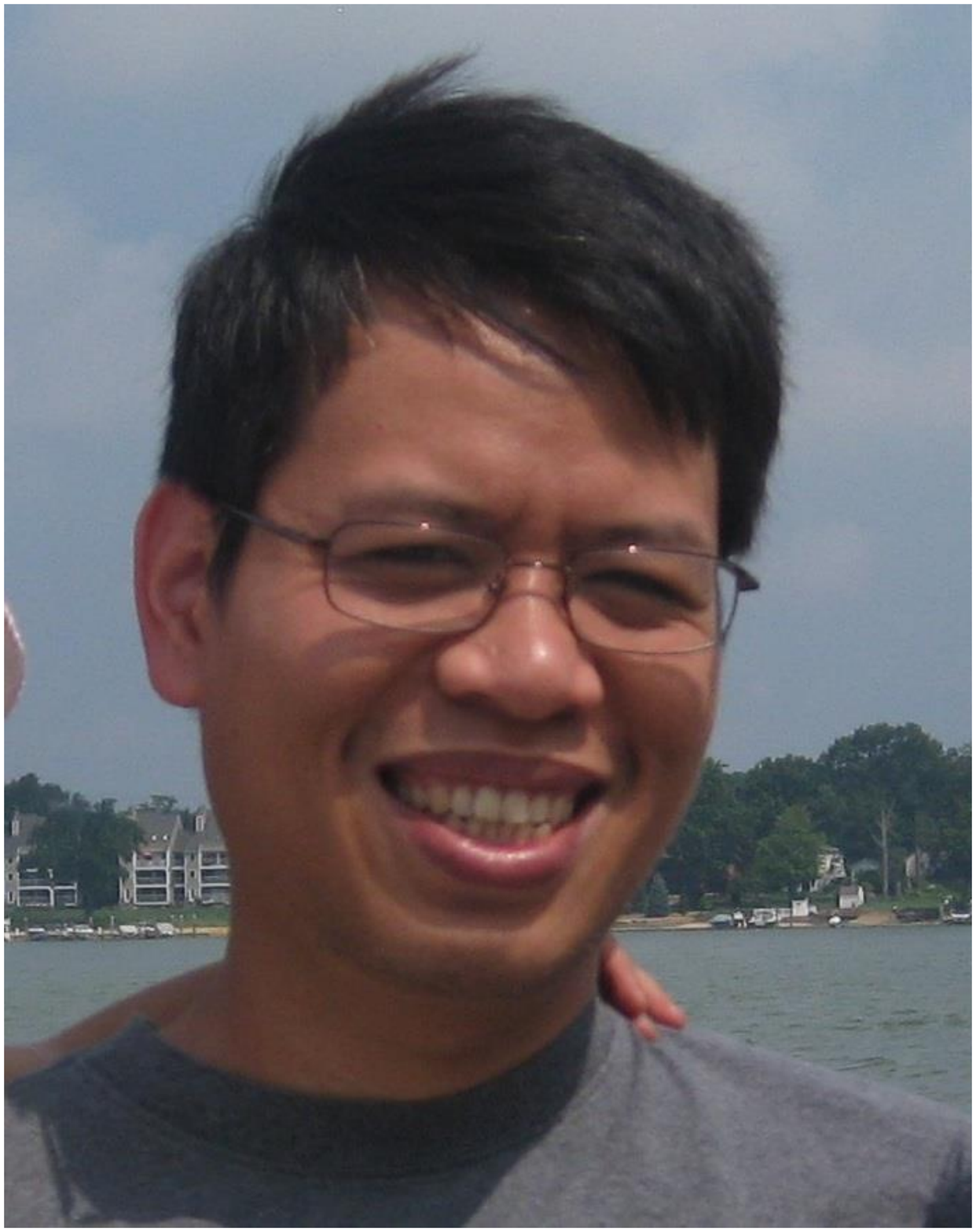}}]{XuanLong Nguyen} received the Ph.D. degree in computer science and the M.S. degree in statistics, both from the University of California, Berkeley. He is currently an Assistant Professor of Statistics at the University of Michigan. His research interests lie in distributed and variational inference, nonparametric Bayesian statistics, and applications to detection/estimation problems in distributed and adaptive systems. Dr. Nguyen is a recipient of the CAREER award from the NSF Division of Mathematical Sciences, the Leon O. Chua Award from the UC Berkeley, the IEEE Signal Processing Society’s Young Author Best Paper award, and an Outstanding Paper award from the International Conference on Machine Learning.
\end{IEEEbiography}
\begin{IEEEbiography}
[{\includegraphics[scale=0.13]{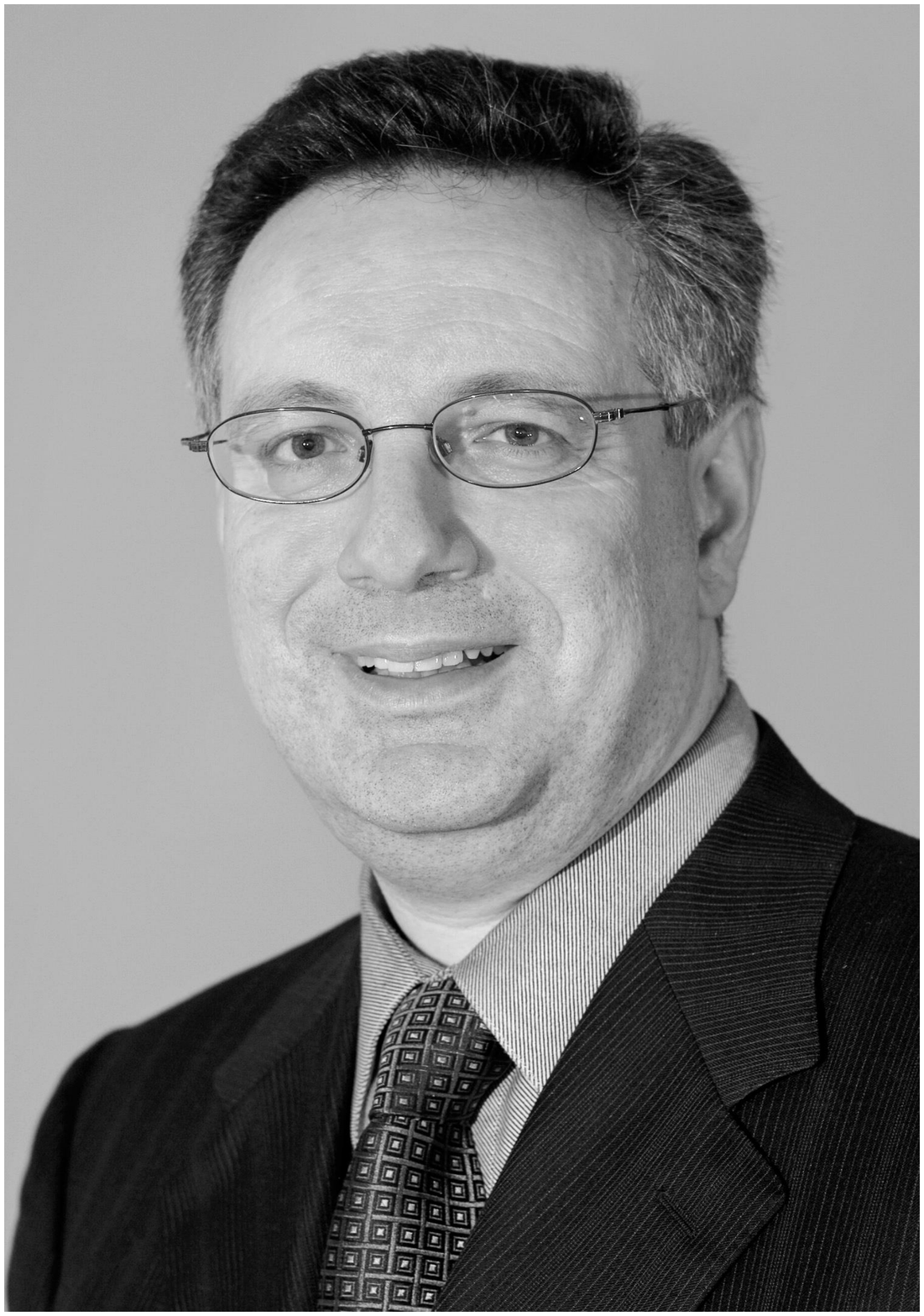}}]{Dennis Assanis} received the Ph.D. degree in Power and Propulsion and the M.S. degrees in Naval Architecture and Marine Engineering and Mechanical Engineering from the Massachusetts Institute of Technology. Dr. Assanis is a Professor in the Department of Mechanical Engineering and is also the Provost, Senior Vice President for Academic Affairs, and Vice President for Brookhaven Affairs at the Stonybrook University, NY, USA. Assanis served as the Jon R. and Beverly S. Holt Professor of Engineering and Arthur F. Thurnau Professor at the University of Michigan, as well as Director of the Michigan Memorial Phoenix Energy Institute, Founding Director of the US-China Clean Energy Research Center for Clean Vehicles and Director of the Walter E. Lay Automotive Laboratory. Dr. Assanis’ research interests lie in the thermal sciences and their applications to energy conversion, power and propulsion, and automotive systems design. His research focuses on analytical and experimental studies of the thermal, fluid and chemical phenomena that occur in internal combustion engines, after-treatment systems, and fuel processors. His efforts to gain new understanding of the basic energy conversion processes have made significant impact in the development of energy and power systems with significantly improved fuel economy and dramatically reduced emissions. His group’s research accomplishments have been published in over 250 articles in journals and international conference proceedings. Dr. Assanis is a Member of the National Academy of Engineering and is an ASME and SAE Fellow.
\end{IEEEbiography}

\end{document}